\documentclass[onecolumn, noshowpacs]{revtex4}
\usepackage{epstopdf}
\usepackage{amssymb}
\usepackage{amsmath}
\usepackage{amsfonts}
\usepackage{graphicx}
\usepackage{color}
\renewcommand{\baselinestretch}{2}
\begin{document}
\renewcommand{\baselinestretch}{2}
\title{An analytical coarse-graining method which preserves the free energy, structural correlations, and thermodynamic state of polymer melts from the atomistic to the mesoscale}
%Variable-level coarse-graining approach to polymer melts: structural and thermodynamic properties.
\author{J. McCarty}
\author{A. J. Clark}
\author{J. Copperman}
\author{M.G. Guenza\footnote{Author to whom correspondence should be addressed. Electronic mail: mguenza@uoregon.edu}}
\affiliation{Department of Chemistry and Institute of Theoretical Science, University of Oregon, Eugene, Oregon 97403}
\date{\today}
\newpage
\begin{abstract}
Structural and thermodynamic consistency of coarse-graining models across multiple length scales is essential for the predictive role of multi-scale modeling and molecular dynamic simulations that use mesoscale descriptions. Our approach is a coarse-grained model based on integral equation theory, which can represent polymer chains at variable levels of chemical details. The model is analytical and depends on molecular and thermodynamic parameters of the system under study, as well as on the direct correlation function in the  $k \rightarrow 0$ limit,  $c_0$. A numerical solution to the PRISM integral equations is used to determine  $c_0$, by adjusting the value of the effective hard sphere diameter, $d_{HS}$, to agree with the predicted equation of state. This single quantity parameterizes the coarse-grained potential, which is used to perform mesoscale simulations that are directly compared with atomistic-level simulations of the same system. We test our coarse-graining formalism by comparing structural correlations, isothermal compressibility, equation of state, Helmholtz and Gibbs free energies, and potential energy and entropy using both united atom and coarse-grained descriptions. We find quantitative agreement between the analytical formalism for the thermodynamic properties, and the results of Molecular Dynamics simulations, independent of the chosen level of representation. In the mesoscale description, the potential energy of the soft-particle interaction becomes a free energy in the coarse-grained coordinates which preserves the excess free energy from an ideal gas across all levels of description. The structural consistency between the united-atom and mesoscale descriptions means the relative entropy between descriptions has been minimized without any variational optimization parameters. The approach is general and applicable to any polymeric system in different thermodynamic conditions.

\end{abstract}
\maketitle
\section{Introduction}
Molecular dynamics (MD) simulations have become instrumental in developing and testing theories in polymer physics since they can provide direct access and physical insight into the time evolution of complex fluids.  However, explicit atom MD simulations are computationally costly which limits their range of applicability to small length and time-scales. This limitation has stimulated the development of numerous coarse-graining methods, which are highly efficient because they represent the system at  a lower resolution, and thus greatly reduce the number of degrees of freedom.\cite{Hall,Dama,Martini,Arun,Kremer,depablo} 
Nonetheless, a major limitation in coarse-graining, which has delayed the widespread use of such models in engineering and material science is the lack of thermodynamic consistency between various representations. For many numerical coarse-graining schemes, there is no guarantee that the resulting behavior of the coarse system will be consistent with what would have been observed by using a more detailed, more-expensive, model.\cite{JOHNSON,LOUIS,FALLER2008} Much of the difficulty in developing consistent coarse-graining approaches stems from the fact that it is not always clear how various many-body effects are incorporated into simple two-body interactions, and how errors in the numerical optimization of a pair potential are propagated to thermodynamic properties.\cite{Anthony2}

In a multiscale procedure, one seeks to link simulations at different levels of description into an ``unified" description of the liquid at all length scales. This requires understanding, in a well-defined and clear way, how to represent a molecule into coarser and coarser units successively, and then to reintroduce the finer structure of the liquid ``a posteriori" in a reverse mapping procedure.\cite{Kremer} This procedure needs to be done with coarse-graining models that ensure the consistency of the structure and the thermodynamics, independent of the degree of coarseness of the molecular model adopted. Conserving both the thermodynamic and dynamical properties while representing the same system with two different degrees of structural refinement, i.e. atomistic and effective unit models, is not trivially done, with the consequence that even the most popular coarse-graining techniques generally produce systems that have the correct structure but too high compressibility and in most cases incorrect free energy. This deficiency can have important consequences when coarse-grained models are used in mesoscale molecular dynamics simulations to study large systems approaching their phase transitions, or the thermodynamic conditions that lead to the possible emergence of kinetically trapped metastable phases, that have geometries desirable for their engineering applications.

In a series of papers\cite{YAPRL,YA2005, SAM2006, SAMB, Anthony, Anthony2}  we have developed a coarse-graining method which is based on statistical mechanical principles, i.e. the Ornstein-Zernike equation and liquid state theory.  The structure of the  integral equation theory warrants the proper calculation of the effective pair interactions between coarse-grained units, which emerge from the propagation through the liquid of the many-body atomistic interactions. Recently, we demonstrated in detail how our coarse-graining procedure is thermodynamically consistent with the atomistic description, within polymer integral equation theory, for a soft-sphere coarse-grained representation where we used the PRISM thread model to parameterize the atomistic model.\cite{McCarty1, PRISM}.  
We also presented a detailed, formal study of the effective potential, when a polymer is represented as a chain of soft blobs with variable size. 
Using the analytical solution for the effective pair potential acting between coarse-grained units, we have demonstrated that the thermodynamic inconsistency frequently observed in numerical coarse-graining schemes arises from a failure to properly consider the long range repulsive tail and related attractive well of the effective potential\cite{AnthonyPRL}. It is our contention that with a reasonable molecular model along with the correct parameters, both structural and thermodynamic properties can be simultaneously preserved in coarse-graining, as it is in our approach, and that this can be achieved without recourse to any numerical re-optimization scheme.  

The purpose of this paper is to implement our variable-level coarse-graining model to derive thermodynamic and structural properties for the variable levels of our coarse-grained description, to investigate consistency and inconsistency of the thermodynamic properties, and to demonstrate the versatility of the approach to model realistic polymer chains. The approach taken here is slightly different from our previous work. In our previous work, we obtained an analytical expression of the effective coarse-grained potential, that depends on the monomer parameter, $c_0=c(k=0)$. Essentially, the parameterization of the potential reduces to finding the single monomer quantity, $c_0$. In this work, we propose to obtain this quantity from the solution to the numerical PRISM equations for a homopolymer semi-flexible chain. This model requires an effective hard sphere diameter, $d_{HS}$ which is chosen such that the pressure calculated from the predicted equation of state agrees with the observed pressures in united atom simulations. Once this parameter is known, the potential is uniquely specified. 

\begin{center}
\begin{figure}
\includegraphics[scale=0.5]{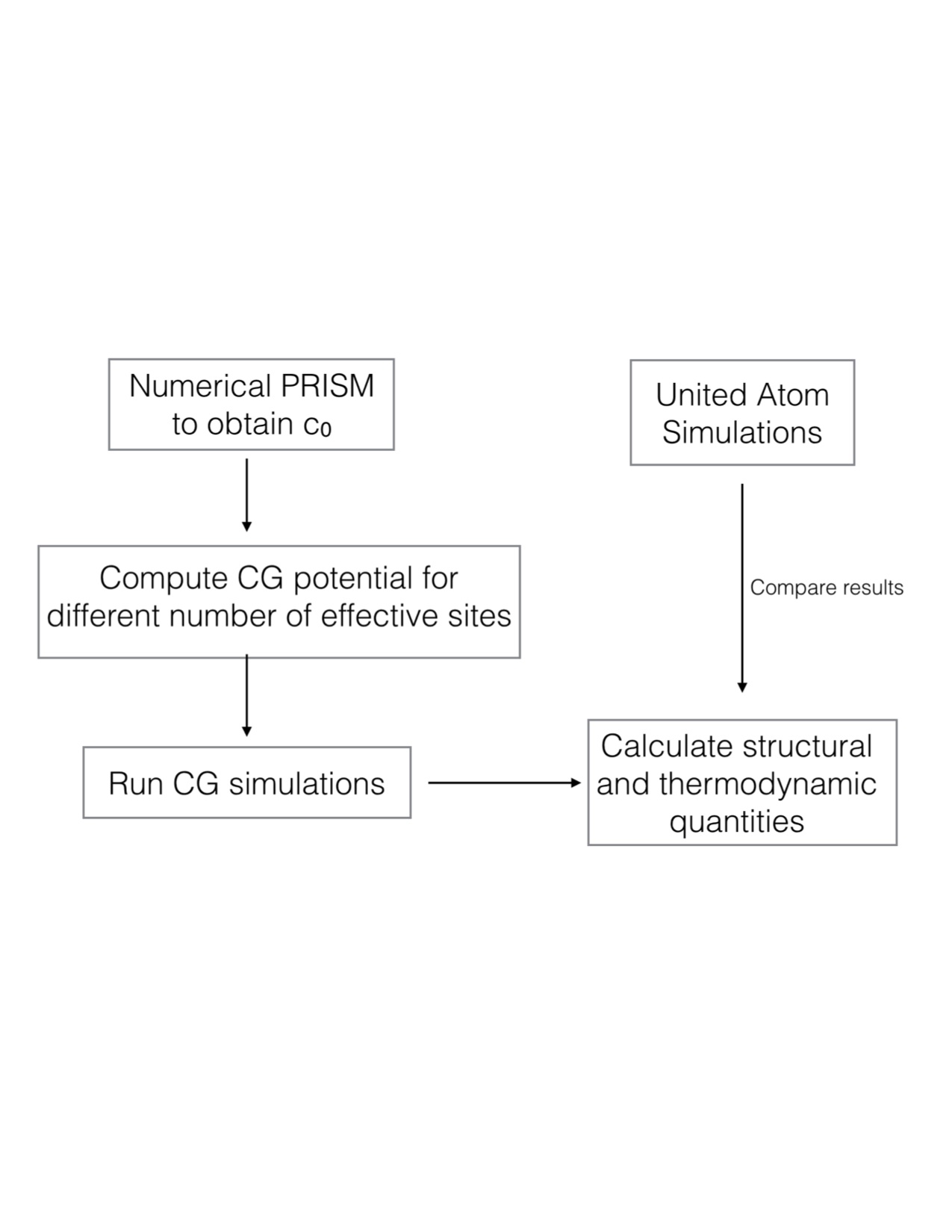}
\caption{Schematic outline of the approach used in this paper. There are two different types of simulations: united atom and coarse-grained. The coarse-grained potentials are obtained by solving the numerical PRISM equation for the parameter $c_0$, which serves as an input into the analytical theory. The potential is then used in coarse-grained simulations with variable numbers of effective sites. Structural and thermodynamic quantities are then compared directly to united atom simulations.}
\label{FG:cgmethod}
\end{figure}
\end{center}

Using the calculated potential we perform simulations of the coarse-grained systems. We then compare thermodynamic quantities and structural quantities of interest from these coarse-grained simulations with united atom simulations. Figure \ref{FG:cgmethod} shows a schematic of the method. There are two sets of simulations: coarse-grained and united atom. The united atom simulations are used only as a test of the method, not too provide information to the coarse-graining approach but to compare thermodynamic properties with the coarse-grained simulations of the same system. Tests are done for a variety of coarse-graining levels, and for systems in different thermodynamic conditions and variable chain length.

Here we consider linear chains of polyethylene melts at various chain lengths and various densities where we represent the chains at different level of chemical detail. We show that our coarse-graining procedure is general and applicable to a wide number of interesting systems. Furthermore, the study presented here shows that our approach is useful in multi-scale simulations of dense polymer systems with specific chemical structure because it is system specific and reproduces the correct equation of state across various levels of coarse-graining.

The paper is divided as follows. We begin Section II with a brief discussion of the background theory that summarizes the procedure to compute the effective potential, in terms of the parameter, $c_0$. In Section III, we introduce the numerical solution of the PRISM approach to calculate the direct correlation function parameter, $c_0$, which is an input parameter to the coarse-grained representation. Then, we summarize the procedure that we follow to perform molecular dynamics simulations of the coarse-grained liquid, and the united atom molecular dynamic simulations that are used to test the theory. In the following sections, the paper presents calculations of thermodynamic properties, specifically structural correlation functions, isothermal compressibility (Section IV), equation of state, Helmholtz and Gibbs free energies, potential energy and entropy (Section V).  For each level of coarse-graining both the analytical formalism for the thermodynamic properties, and the numerical solution of the thermodynamics using mesoscale simulations of the coarse-grained system, as well as the calculation of the thermodynamic properties by united atom molecular dynamic simulations are presented.  The purpose of this cross testing using multiple approaches is to demonstrate, without any ambiguity, the consistency (or inconsistency) of the thermodynamic properties for the variable-level coarse-graining model, independent of the chosen level of coarse-graining.  The study confirms the agreement of the structure and thermodynamics of the coarse-grained representation, with the united atom description of the same liquid through traditional MD simulations, and validates the analytical description of the thermodynamic properties presented in this paper.

\begin{center}
\begin{figure}
\includegraphics[scale=0.5]{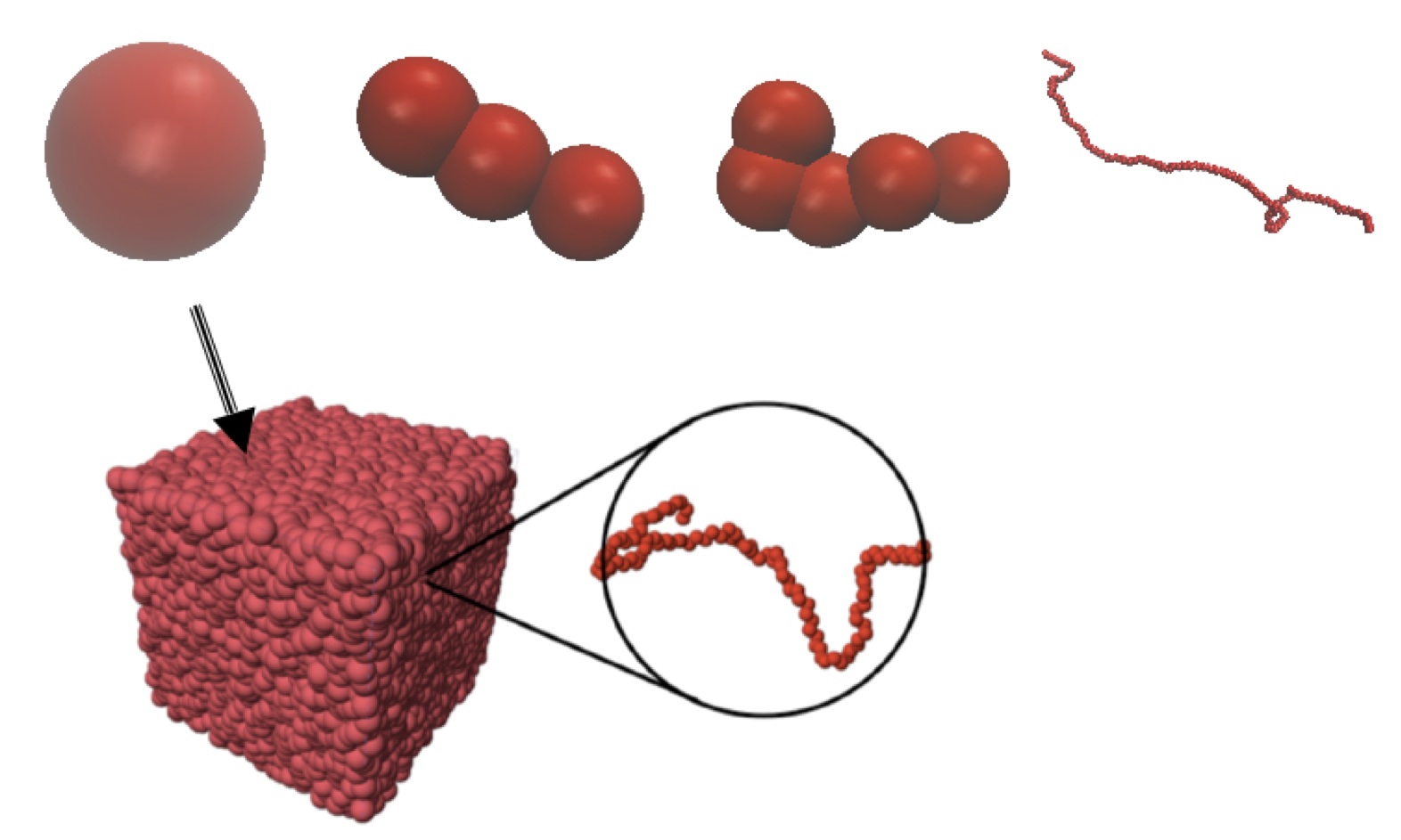}
\caption{Coarse-graining at multiple block-levels. The far right is a snapshot of a typical configuration from an United Atom Molecular Dynamics simulation of polyethylene. Moving from right to left the same chain is represented by 5 blocks, 3 blocks, to a single soft sphere. Below is a snapshot of a typical mesoscale simulations, where each sphere represents a single polyethylene chain.}
\label{FG:snapshot}
\end{figure}
\end{center}

\section{Background Information: Variable-level Coarse-Grained Description of Polymer Melts}
\label{Coarse Grained Potential}
Our course grained description represents each polymer chain in a melt as a single soft sphere\cite{YAPRL,YA2005,SAM2006} or as a collection of $n_b$ soft connected blobs\cite{SAMB, Anthony, Anthony2}.  
While our previous papers provided a theoretical derivation of the variable-level coarse-grained representation, here the numerical procedure is only summarized for completeness. In a series of snapshots Figure \ref{FG:snapshot} shows how a single polymer chain is represented first by several spheres, and then by a single soft sphere, allowing for the simulation of many more polymers than would otherwise be possible to simulate.

 We take the polymer to be comprised of equivalent sites of $N$ monomers with a chain density $\rho_{ch}$, and an effective segment length $\sigma=\sqrt{6/N}R_g$ with $R_g$ being the polymer radius of gyration and $R_{gb}=R_{g}/\sqrt{n_b}$ the blob radius of gyration. The number of underlying monomers per blob is given as $N_b=N/n_b$. Coarse-grained or fictitious interacting sites are taken to be located at the center of mass of the polymer chain for the soft sphere model or at the centers of mass of several monomers along the same chain for the connected blob model. The relation between center of mass fictitious sites and real monomer sites, originally proposed by Krakoviack \emph{et al}\cite{Krakoviack} is derived by solving a generalized matrix Ornstein Zernike equation, and was extended by Clark  and Guenza \cite{Anthony} for the multi-blob model. We now summarize the model for the case of one, three, and multiple effective sites (see Figure \ref{FG:snapshot}).

\subsection{Soft Sphere Potential}
 For the case where the number of blobs is one, meaning each chain is represented as a single interaction site, one recovers the relation,
\\
\begin{equation}
\hat{h}^{cc}(k)=\left[\frac{\hat{\omega}^{cm}(k)}{\hat{\omega}^{mm}(k)}\right]^2 \hat{h}^{mm}(k),
\label{EQ:HCCK5}
\end{equation}
where $\hat{h}^{cc}(k)$ is the total correlation function between center of mass, fictitious sites. $\hat{h}^{mm}(k)$ is the total correlation function between monomers, and $\hat{\omega}^{mm}(k)$ and $\hat{\omega}^{cm}(k)$ are the intramolecular distributions between monomers or between monomers about the center of mass respectively. The monomer total correlation function, $\hat{h}^{mm}(k)$, is given in Fourier space by the PRISM integral equation for a homopolymer fluid\cite{PRISM1,PRISM2},
\begin{equation}
\hat{h}^{mm}(k)=\hat{\omega}^{mm}(k)\hat{c}^{mm}(k)[\hat{\omega}^{mm}(k)+\rho \hat{h}^{mm}(k)],
\label{EQ:OZch5}
\end{equation}
where $\rho=N\rho_{ch}$ is the number density of monomer sites, and $\hat{c}^{mm}(k)$ is the monomer direct correlation function. As an approximation, valid at large r, we take the direct correlation function to be a constant in Fourier space, $\hat{c}^{mm}(k)=\hat{c}^{mm}(0)=c_0$. Substitution of Equation \ref{EQ:OZch5} into Equation \ref{EQ:HCCK5} gives,
\begin{equation}
\hat{h}^{cc}(k)=\frac{c_0[\hat{\omega}^{cm}(k)]^2}{1-\rho c_0 \hat{\omega}^{mm}(k)} \ .
\label{EQ:CLARK5}
\end{equation}
For the soft sphere model, the direct correlation function between coarse-grained sites, $\hat{c}^{cc}(k)$, is identical to that for a simple fluid,
\begin{equation}
\hat{c}^{cc}(k)=\frac{\hat{h}^{cc}(k)}{1+\rho_{ch}\hat{h}^{cc}(k)} \ .
\label{EQ:dircorrk}
\end{equation}
The effective soft sphere potential is given by the hypernetted chain (HNC) closure\cite{HansenMcDonald},
\begin{equation} 
\frac{v^{cc}(r)}{k_BT}=-\ln{[h^{cc}(r)+1]}+h^{cc}(r)-c^{cc}(r) \ .
\label{EQ:Closure2}
\end{equation}
The monomer intramolecular distribution is given by the Debye formula,
\begin{equation}
\omega^{mm}(k)=\frac{2 N (e^{-k^2 R_g^2}+k^2 R_g^2-1)}{k^4 R_g^4} \ ,
\label{EQ:Debye}
\end{equation}
whereas the monomer-cm distribution is given by 
\begin{equation}
\omega^{cm}(k)=\frac{N\sqrt{\pi}}{kR_g}e^{-k^2R_g^2/12}\text{erf} \left[\frac{kR_g}{2}\right] \ .
\label{EQ:WCMK}
\end{equation}
The analytical solution of $v^{cc}(r)$ for polymer melts, i.e. high density mean-field solution, has been presented in a recent publication, together with the study of the scaling behavior.\cite{Anthony2} In this paper we calculate the potential numerically. The procedure formulated in this manner is easy to implement. Using the definitions of the intramolecular and intermolecular distribution functions, $\hat{h}^{cc}(k)$ is simply calculated and its  subsequent numerical Fourier gives the effective potential from Equation \ref{EQ:Closure2}. This potential is used as an input to Molecular Dynamics simulations of soft spheres. Note that the formalism only depends on molecular ($N$ and $R_g$) and thermodynamic parameters ($T$, $\rho$, and $\rho_{ch}$), as well as on the system compressibility through the direct correlation function at $k=0$, $c_0$.

\subsection{Tri-Block Potential}

For chains represented by three coarse grained sites per chain (tri-block model) the two terminal coarse grained sites are distinct from the central coarse grained site giving rise to three distinct block combinations of block-block total correlation functions. Using the notation from our previous work\cite{Anthony} the intermolecular total correlation functions between block centers, $\hat{h}^{bb}_{\alpha \beta}(k)$ with block indices $\alpha$ and $\beta$, are given as a function of the new omega distributions for the tri-block structure as 
\begin{eqnarray}
\hat{h}^{bb}_{11}(k)&=&\hat{h}^{bb}_{13}(k)=\hat{h}^{bb}_{31}(k)=\hat{h}^{bb}_{33}(k)  \ , \nonumber \\
&=&\frac{c_0[\hat{\omega}^{bm}_0(k)+\hat{\omega}^{bm}_1(k)+\hat{\omega}^{bm}_2(k) ]^2}{1-\rho c_0 \hat{\omega}^{mm}(k)} \ ,
\label{EQ:hbb11}
\end{eqnarray}

\begin{eqnarray}
\hat{h}^{bb}_{12}(k)&=&\hat{h}^{bb}_{23}(k)=\hat{h}^{bb}_{21}(k)=\hat{h}^{bb}_{32}(k) \ , \nonumber \\
&=&\frac{c_0[(\hat{\omega}^{bm}_0(k)+\hat{\omega}^{bm}_1(k)+\hat{\omega}^{bm}_2(k))(\hat{\omega}^{bm}_0(k)+2\hat{\omega}^{bm}_1(k))]^2}{1-\rho c_0 \hat{\omega}^{mm}(k)} \ ,
\label{EQ:hbb12}
\end{eqnarray}
and
\begin{equation}
\hat{h}^{bb}_{22}(k)=\frac{c_0[\hat{\omega}^{bm}_0(k)+2\hat{\omega}^{bm}_1(k)]^2}{1-\rho c_0 \hat{\omega}^{mm}(k)} \ .
\label{EQ:hbb22}
\end{equation}

The direct correlation functions are then expressed in term of the intramolecular distributions by inserting Equations \ref{EQ:hbb11}-\ref{EQ:hbb22} into the matrix Ornstein-Zernike equation of the coarse-grained fluid. The resulting direct correlation functions are lengthy but presented here in their full form,
\begin{align}
\hat{c}^{bb}_{11}(k)&=-\frac{1}{\hat{D}(k)}\left(\hat{h}^{bb}_{11}(k)-2\hat{h}^{bb}_{12}(k)\hat{\omega}^{bb}_{1}(k) \right. \nonumber \\
& \left. +\hat{h}^{bb}_{22}(k)(\hat{\omega}^{bb}_1(k))^2-\rho_{ch}(\hat{h}^{bb}_{12}(k))^2 +\rho_{ch}\hat{h}^{bb}_{11}(k)\hat{h}^{bb}_{22}(k) \right)  \ ,
\label{EQ:cbbk11}
\end{align}

\begin{align}
\hat{c}^{bb}_{12}(k)&=-\frac{1}{\hat{D}(k)}\left(\hat{h}^{bb}_{12}(k)(1+2(\hat{\omega}^{bb}_1(k))^2+\hat{\omega}^{bb}_2(k))+2\rho_{ch}(\hat{h}^{bb}_{12}(k))^2\hat{\omega}^{bb}_1(k) \right. \nonumber \\
& \left. -\hat{\omega}^{bb}_1(k)(2\hat{h}^{bb}_{11}(k)+\hat{h}^{bb}_{22}(k)+\hat{h}^{bb}_{22}(k)\hat{\omega}^{bb}_2(k)+2\rho_{ch}\hat{h}^{bb}_{22}(k)\hat{h}^{bb}_{11}(k))\right) \ ,
\label{EQ:cbbk12}
\end{align}

\begin{align}
\hat{c}^{bb}_{22}(k)&=-\frac{1}{\hat{D}(k)}\left(4(\hat{\omega}^{bb}_1(k))^2\hat{h}^{bb}_{11}(k)-4\hat{\omega}^{bb}_1(k)(1+\hat{\omega}^{bb}_2(k))\hat{h}^{bb}_{12}(k) \right. \nonumber \\
& \left.-2\rho_{ch}(1+\hat{\omega}^{bb}_2(k))(\hat{h}^{bb}_{12}(k))^2+\hat{h}^{bb}_{22}(k)(1+\hat{\omega}^{bb}_2(k))(1+\hat{\omega}^{bb}_2(k)+2\rho_{ch}\hat{h}^{bb}_{11}(k))\right) \ ,
\label{EQ:cbbk22}
\end{align}
with 
\begin{align}
\hat{D}(k)&=(2(\hat{\omega}^{bb}_1(k))^2-\hat{\omega}^{bb}_2(k)-1)(1-2(\hat{\omega}^{bb}_1(k))^2+\hat{\omega}^{bb}_2(k)-4\rho_{ch}\hat{\omega}^{bb}_1(k)\hat{h}^{bb}_{12}(k)+ \nonumber \\
& \rho_{ch}\hat{\omega}^{bb}_2(k)\hat{h}^{bb}_{22}(k)+\rho_{ch}(2\hat{h}^{bb}_{11}(k)+\hat{h}^{bb}_{22}(k)-2\rho_{ch}(\hat{h}^{bb}_{12}(k))^2+2\rho_{ch}\hat{h}^{bb}_{11}(k)\hat{h}^{bb}_{22}(k))) \ .
\label{EQ:Dtot}
\end{align}

In order to to obtain the tri-block potential, we use the results for the tri-block intramolecular distributions calculated by Clark and Guenza\cite{Anthony2}. The monomer-monomer distribution is given by the generalized Debye formula,
\begin{equation}
\hat{\omega}^{mm}(k)=\frac{2N}{n_b^2k^4R_{gb}^4}(k^2R_{gb}^2n_b-1+e^{-n_bk^2R_{gb}^2}) \ .
\label{EQ:wmmkav}
\end{equation}
For the block-monomer and block-block distributions, distinct block sites are represented by Greek indices $\alpha$ and $\beta$ and the block separation is given by $\gamma=|\alpha -\beta|$. For the tri-block model $\gamma \in \{0,1,2\} $ and the distributions of monomers about blocks on the same site, $\gamma=0$, is given as
\begin{equation}
\omega^{bm}_0(k)=\frac{N\sqrt{\pi}}{kR_g}e^{-k^2R_g^2/12}\text{erf} \left[\frac{kR_g}{2}\right] \ ,
\label{EQ:WBMK0}
\end{equation}
whereas the block-monomer distribution of monomers on block, $\alpha$, with block $\beta$ is 
\begin{equation}
\omega^{bm}_{\gamma \neq 0}(k)=\frac{N}{k^2R_{gb}^2}   e^{-k^2R_{gb}^2[2+6(\gamma-1)]/6}   (1-e^{-k^2R_{gb}^2}) \ .
\label{EQ:WBMkg}
\end{equation}
Lastly, the block-block distribution is given as
\begin{equation}
\omega^{bb}_\gamma(k)=Ne^{-k^2R_{gb}^2[4+6(\gamma-1)]/6} \ .
\label{EQ:WBBkg}
\end{equation}
Once again Equations \ref{EQ:wmmkav}-\ref{EQ:WBBkg} are used to derive expressions for $h^{bb}(k)$ and $c^{bb}(k)$, using Equations \ref{EQ:hbb11}-\ref{EQ:hbb22} and Equations \ref{EQ:cbbk11}-\ref{EQ:cbbk22}. The potential is then calculated numerically after Fourier transform using the HNC closure relation, yielding three potentials  between block types, $v_{AA}(r)$, $v_{BB}(r)$, and $v_{AB}(r )$, depending if the block is inside the polymer or at its ends.\cite{Anthony2}

The formalism just presented to describe position-specific correlation functions for tri-block renormalization can be extended to treat five-block renormalization.\cite{Anthony2} The distribution functions are formally complex, but the procedure of deriving them is straightforward. 

\subsection{Block Averaged Potential}
Although the expressions for site specific block properties are complicated, when the number of blocks is large,  the formalism can be drastically simplified. Specifically we have observed that when there are more than five coarse grained sites per chain, end effects become negligible and it is possible to use a block-averaged description which simplifies the expressions. A detailed analysis of the properties of the block-averaged potential is presented in a recent paper together with the analytical solution of the expression for the potential and the analysis of the scaling exponents.\cite{Anthony2}. 

We use here a numerical procedure to calculate the characteristic distribution functions and the potentials, which enter the molecular dynamic simulations. The numerical procedure begins the same way by calculating the intramolecular distributions in the block averaged limit. To maintain consistency with the expressions in our previous work\cite{Anthony2}, we introduce the normalized intramolecular distribution $\Omega(k)=\omega(k)/N$, for the block-block (bb), block-monomer (bm), and monomer-monomer (mm) distributions.  The normalized block-monomer and the block-block distributions in the chain averaged limit are
\begin{equation}
\hat{\Omega}^{bm}(k) = \frac{1}{n_b} \left[ \frac{\sqrt{\pi}}{k R_{gb}}Erf\left( \frac{kR_{gb}}{2} \right)e^{-\frac{k^2R_{gb}^2}{12}} - 2 \left( \frac{e^{-n_b k^2R_{gb}^2}-n_b e^{-k^2R_{gb}^2}+n_b-1}{k^2 R_{gb}^2 n_b (e^{-k^2 R_{gb}^2}-1)} \right) e^{-k^2R_{gb}^2/3} \right] \ ,
\label{EQ:omegabavbmch5}
\end{equation}
and 
\begin{equation}
 \hat{\Omega}^{bb}(k) = \frac{1}{n_b} + 2\left[\frac{e^{-n_b k^2R_{gb}^2}-n_b e^{-k^2R_{gb}^2}+(n_b-1)}{n_b^2(e^{-k^2R_{gb}^2}-1)^2} \right] e^{-2k^2R_{gb}^2/3} \ .
 \label{EQ:Ombbkavch5}
\end{equation}
The monomer distribution $\Omega^{mm}(k)$ is again given by equation \ref{EQ:wmmkav} normalized by $N$. 

The next step is to use the result from solving the matrix Ornstein-Zernike equation for $h^{bb}(k)$,
\begin{equation}
\hat{h}^{bb}(k)=\left[\frac{\hat{\Omega}^{bm}(k)}{\hat{\Omega}^{mm}(k)}\right]^2 \hat{h}^{mm}(k),
\label{EQ:HbbK5}
\end{equation}
and, from the Ornstein-Zernike relation 
%, which is done numerically using tabulated k-values,
\begin{equation}
 \hat{h}^{bb}(k)= n_b \hat{\Omega}^{bb}(k) \hat{c}^{bb}(k) \left[ n_b \hat{\Omega}^{bb}(k) +\rho_b \hat{h}^{bb}(k) \right]\ ,
 \label{EQ:hbbkavch5}
\end{equation}
in the direct correlation function, calculated as
\begin{equation}
\hat{c}^{bb}(k)=\frac{\hat{h}^{bb}(k)}{n_b \hat{\Omega}^{bb}(k)\left[n_b \hat{\Omega}^{bb}(k)+\rho_b \hat{h}^{bb}(k)\right]} \ .
\label{EQ:cbbkavch5}
\end{equation}

While the analytical solution of the potential has been presented earlier,\cite{Anthony2}
in this paper we solve the potential numerically. 
First, the direct and total distribution functions in real space are calculated from numerical Fourier transform of Equation \ref{EQ:hbbkavch5} and Equation \ref{EQ:cbbkavch5}. Then, the interaction potential is calculated numerically evaluating the HNC potential, as above,
\begin{equation} 
\frac{v^{bb}(r )}{k_BT}=-\ln{[h^{bb}(r )+1]}+h^{bb}(r )-c^{bb}(r ) \ .
\label{EQ:Closure3}
\end{equation}
% ~~~~~~~~~~~~~~~~~~~~~~~~~~~~~~~~~~~~~~~~~~~~

%~~~~~~~~~~~~~~~~~~~~~~~~~~~~~~~~~~~~~~~~~~~~~~
\section{Methods}

\subsection{Numerical PRISM Solution for the Calculation of the Direct Correlation Function}
The only non-trivial parameter in our formalism is the direct correlation function in the large scale limit.\cite{McCarty1} For a realistic representation of the polymer chain it is useful to work with the numerical solution of the PRISM equations, which are solved to obtain the monomer direct correlation parameter, $c_0$, which is then used as an input in the expressions of the coarse-grained potentials. This is the only non-trivial parameter that enters our approach, i.e. the only parameter that does not, trivially, describe physical or molecular quantities. Other parameters, besides the direct correlation $c_0$, are the thermodynamic properties of temperature, $T$, and density, $\rho$, as well as the structural properties of $N$, the number of monomers, and the effective segment size, $\sigma$. These parameters allow the method to be readily applied to a variety of polymers in variable experimental conditions. 

An analytical definition of the $c_0$ parameter was obtained using the Gaussian thread model, which relies on the description of the polymer chain as infinitely long and infinitely thin, and is the model used to represent polymers in field theory.\cite{McCarty1}. While the PRISM thread model represents an idealized limiting case, it is not expected to give quantitative predictions for real chains of finite length and thickness. In this paper we use a self-consistent numerical solution of the PRISM  equation.

\begin{center}
\begin{figure}
\includegraphics[scale=0.5]{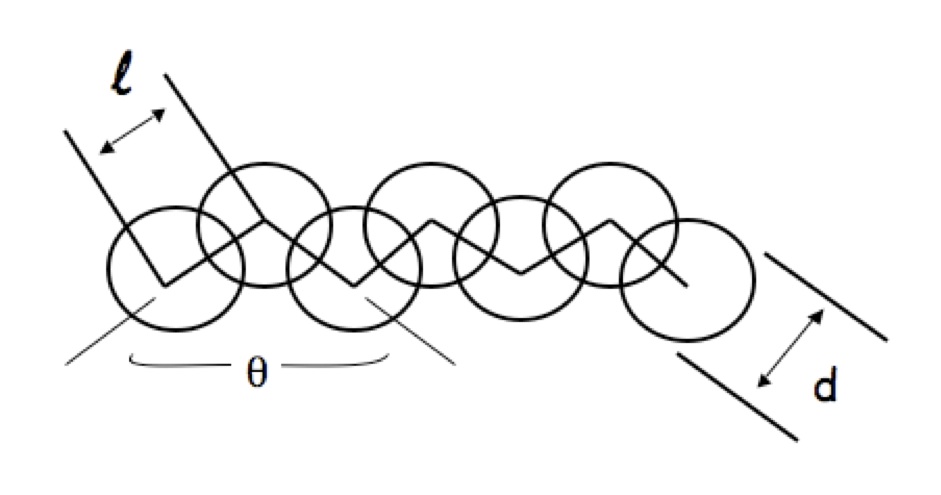}
\caption{Semiflexible chain model used as the molecular model for numerical PRISM equations. This model allows for a realistic estimate of $c_0$, which is input into the coarse-grained model.}
\label{FG:tangenths}
\end{figure}
\end{center}

The monomer total correlation function, $\hat{h}^{mm}(k)$, is given in Fourier space by the PRISM integral equation for a homopolymer fluid\cite{PRISM1,PRISM2},
\begin{equation}
\hat{h}^{mm}(k)=\hat{\omega}^{mm}(k)\hat{c}^{mm}(k)[\hat{\omega}^{mm}(k)+\rho \hat{h}^{mm}(k)],
\label{EQ:OZappend}
\end{equation}
where $\rho=N\rho_{ch}$ is the number density of monomer sites, and $\hat{c}^{mm}(k)$ is the monomer direct correlation function. We adopt the semiflexible chain model\cite{HonnellCurro} as depicted in Figure \ref{FG:tangenths}. In its simplest form, the model requires three parameters, the monomer hard sphere diameter, $d_{HS}$, the bond length, $l$, and the bond angle, $\theta$. For polyethylene chains we use the values of $l=1.54$ and $\theta=141.7$ to agree with the stiffness $q=-<\cos{\theta}>=0.785$, which has been shown to be reasonable for linear polyethylene models. To estimate the intramolecular distribution, we then use the Koyama distribution which can be calculated using the method from reference \cite{HonnellCurro}.

Having the intramolecular distribution, we solve the PRISM equation, Equation \ref{EQ:OZch5}, with the reference molecular mean spherical approximation (R-MMSA) closure given by\cite{Yethiraj1,Yethiraj2}
 \begin{eqnarray}
 \omega(r)*c(r)*\omega(r)&=\omega(r)*c^{(0)}(r)*\omega(r)-\omega(r)*\beta v(r)*\omega(r),  \nonumber \ \
 &r> d_{HS},                                               
 \label{EQ:RMMSA}
 \end{eqnarray}
where the asterisks denote convolution integrals, $c^{(0)}$ is the direct correlation function of a reference purely hard sphere chain, and $v(r)$ is the attractive part of the potential given as a Lennard Jones tail, 
\begin{eqnarray}
v(r)&=4\epsilon \left[\left(\frac{\sigma}{r}\right)^{12}-\left(\frac{\sigma}{r}\right)^{6}\right], \ \ \ r> \sigma \nonumber \ \
&v(r)=0, \ \ \ \ r< \sigma
\label{EQ:VR}
\end{eqnarray}
Inside the hard core, impenetrability is ensured by enforcing the condition
\begin{equation}
h(r)=-1, \ \ \ r< d_{HS}.
\label{EQ:HardCore}
\end{equation}

Once the closure is specified by Equations \ref{EQ:RMMSA} - \ref{EQ:HardCore}, the PRISM equation is solved numerically using the standard Picard iteration with fast Fourier transform or with the KINSOL nonlinear algebraic solver available through the SUNDIALS suite of programs.\cite{sundials}

Once a solution of is achieved, the value of $c_0$ is determined as 
\begin{equation}
c_0=\frac{\hat{h}(0)}{\rho N\hat{h}(0) +N^2} \ .
\label{EQ:co}
\end{equation}

\begin{figure}
\includegraphics[scale=0.6]{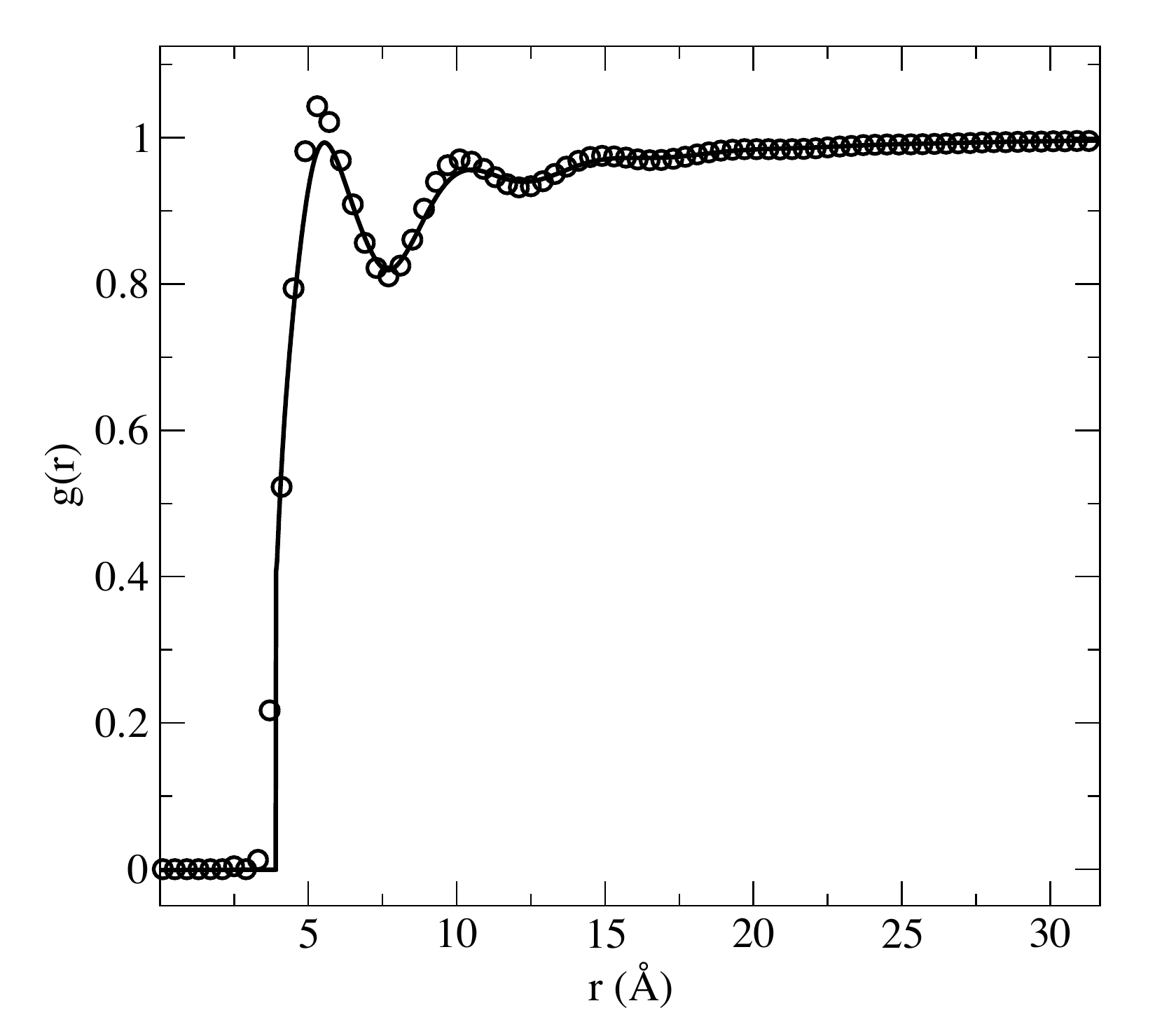}
\caption{The monomer radial distribution function for PE with $N=100$ at a monomer density $\rho=0.03355$ sites/$\AA^3$ calculated from United Atom simulations (circles) and from PRISM calculation (solid line) using the SFC model with $d_{HS}=3.9 \AA$ which gives the best agreement between the pressures}.
\label{FG:grplot}
\end{figure}

\begin{figure}
\includegraphics[scale=0.6]{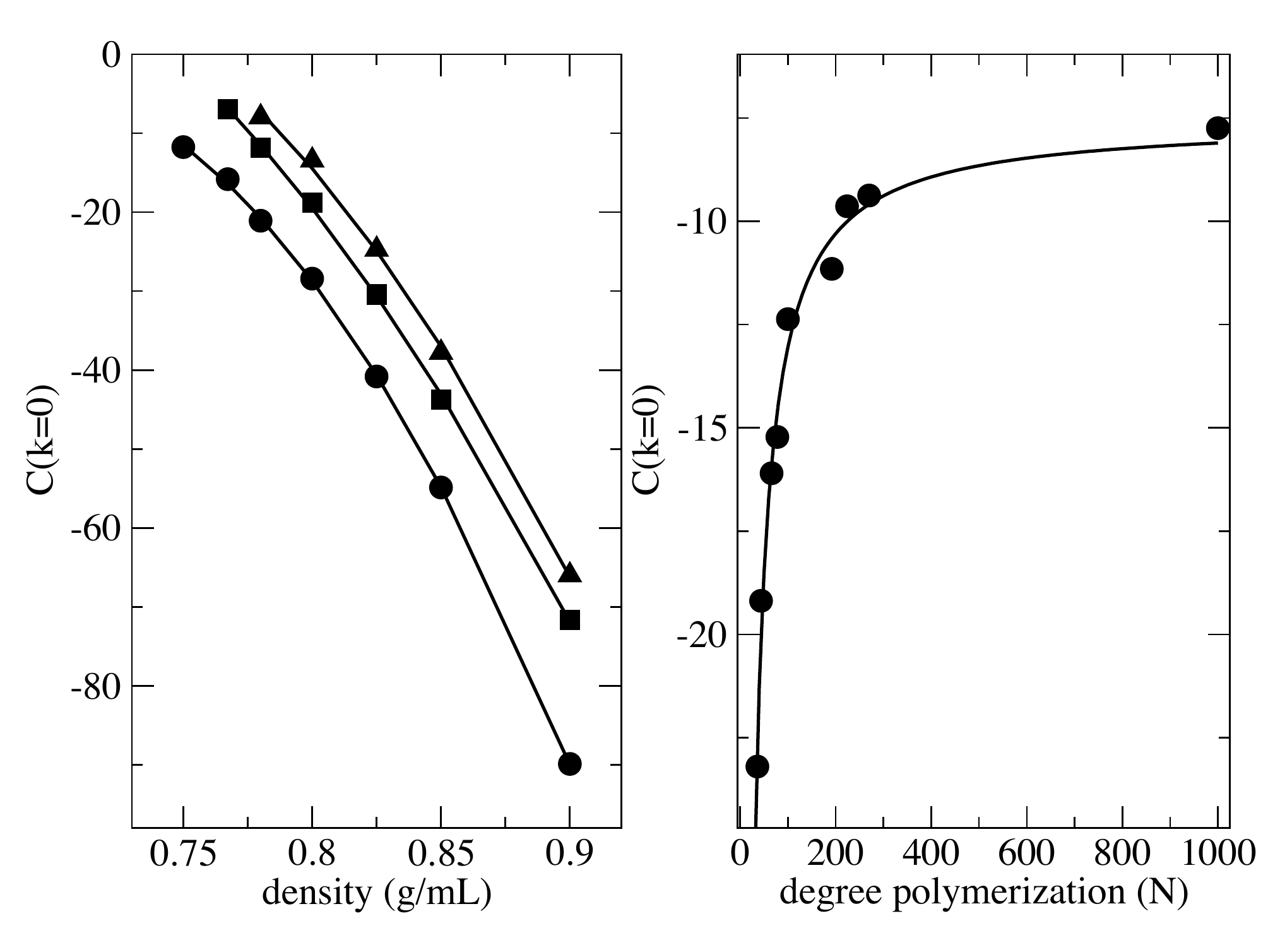}
\caption{Left: Calculated value of $c_0$ from numerical PRISM using a simple semi flexible chain chain model as a function of the monomer site density. Three different chain lengths are shown; $N=44$ (circles), $N=100$ (squares) and $N=200$ (triangles). The line is a fit to a quadratic polynomial to serve as a guide to the eye. Right: Calculated values of $c_0$ from numerical PRISM as a function of polymer chain length at fixed density, $\rho=0.03153$ sites/$\AA^3$ (circles). The solid line is a fit to the form $a+b/N$.}
\label{FG:FigCo}
\end{figure}

In performing numerical PRISM calculations we adjust the hard-sphere diameter, $d_{HS}$, to agree with the predicted equation of state given by Equation \ref{EQ:Soln1ch4}, so that the pressure agrees with United Atom simulation data and $c_0$ determined from Equation \ref{EQ:co}. We also check that this value of $d_{HS}$ yields a good representation of the monomer structure as given by the radial distribution function, $g(r)$. Figure \ref{FG:grplot} shows $g(r)$ calculated from PRISM theory and from United Atom simulations for $N=100$ and a monomer density if $\rho=0.03355$ sites/$\AA^3$.
Figure \ref{FG:FigCo} shows a plot of $c_0$ for different densities for a chain of $N=44$, $N=100$, and $N=200$. Most noticeably, the dependence of $c_0$ on density is stronger than the linear dependence predicted by the idealized thread model. The right side of Figure \ref{FG:FigCo} shows the $c_0$ as a function of chain length for $\rho=0.03153$ sites/$\AA^3$. The value of $c_0$ approaches an asymptotic value for large $N$ which corresponds to a leveling off of the chain-length dependence of the pressure. This limiting value of $c_0$ can be used to perform coarse-grained simulations of large chains where all-atom simulations are prohibitive.

\begin{figure}
\includegraphics[scale=0.5]{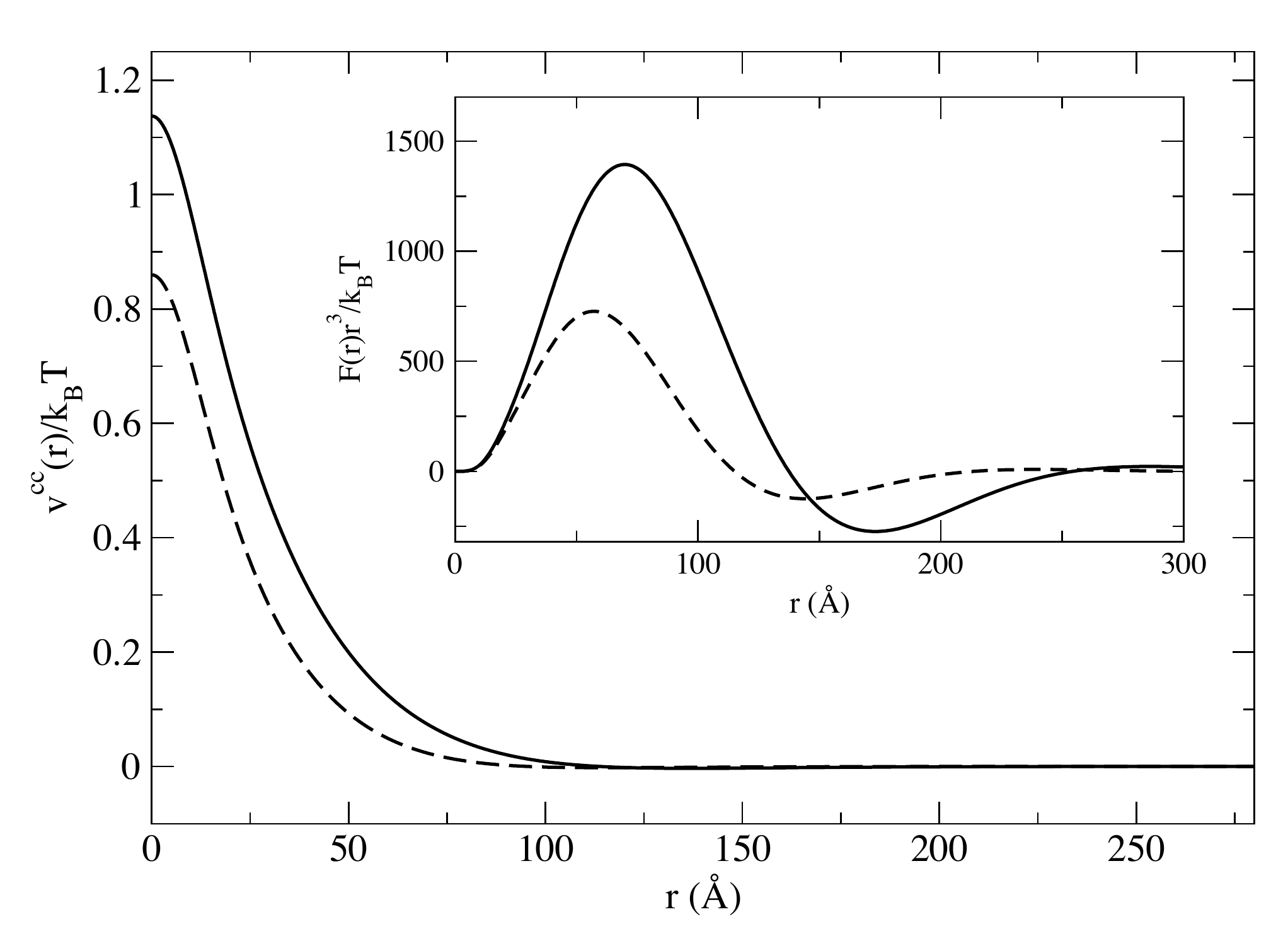}
\caption{The effective pair potential for PE100 at an intermediate density of $\rho=0.03656$ sites/$\AA^3$ The solid line is the potential calculated using numerical PRISM with the SFC chain model. The dashed line is the same potential using the thread model value for $c_0$ for comparison. The inset shows the virial force.}
\label{FG:Virial}
\end{figure}

It is interesting to compare the effective pair potential obtained from the PRISM calculation using the semi flexible chain model (SFC) with that obtained from the thread model. Figure \ref{FG:Virial} shows the effective pair potential for the intermediate density, $\rho=0.03656$ sites/$\AA^3$, where the molecular model input into the theory is calculated from PRISM theory for both the semi flexible chain model and the thread model. The SFC model results in a stronger and longer-ranged repulsive core; however the attractive well is small and of comparable magnitude. The inset shows the virial force $F(r)r^3/k_BT$ which is used to calculate the pressure equation. The repulsive contribution is much larger for the case of the semi flexible chain model. This is due to the finite-sized hard sphere core in the semi flexible chain model, which results in a stronger repulsive force. Because the radial distribution function is fairly insensitive to the shape of the potential, the structure is nearly identical between the thread model and the semi flexible chain model; however, because the pressure depends  on the virial force, and it is extremely sensitive to the potential, even small deviation in the shape of the force can lead to different compressibility, and thus the semi flexible chain model should be used since the value of $c_0$ chosen is such as to match the pressure from the united-atom simulations. 

\subsection{Mesoscale Molecular Dynamics Simulations} 
In our \textit{coarse-grained models} each polymer chain is represented as a collection of soft spheres, where each sphere includes a large section of the polymer chain, which is larger than the polymer persistence length.  As a limiting case, we represent the whole chain with a single, soft interaction site, i.e. the soft-sphere model. For this level of coarse-graining, the relevant information is on length-scales larger than a single polymer chain, and the model affords the largest computational gain.

In general, we call simulations performed at these various multi-block levels of description Mesoscale (MS) simulations, since they represent the system at a level of description in between an all-atom and a continuum level.  Having a description that systematically represents the chain at variable levels of detail, enables us to connect information in a multi-scale procedure.\cite{Jaymultiscale} 
Mesoscale simulations of point particles allow for the simulation of many more particles than the united-atom simulations. The increased particle number makes these simulations suitable to study bulk, large-scale properties such as shearing or phase transitions. 

The soft particles interact via the coarse-grained potentials described in the previous section. 
For the multi-block model, bonded sites are given a bond potential derived from the direct Boltzmann inversion of the probability distribution of the effective bond length,
\begin{equation}
v_{\text{bond}}(r)=-k_BT\ln{\left[P(r)/r^2\right]} 
\label{EQ:Bondch5}
\end{equation}
where 
\begin{equation}
P(r)=4\pi \left[\frac{3}{\pi 8 R_{gb}^2}\right]^{3/2}r^2 \text{exp}\left[-\frac{3r^2}{8 R_{gb}^2}\right].
\label{EQ:BondGaussch5}
\end{equation}
%\begin{equation}
%v_{bond}=\frac{3 k_BT r^2}{8 R_{gb}^2}.
%\end{equation}

The angle potential is similarly computed
\begin{equation}
v_{\text{angle}}(\theta)=-k_BT \ln{\left[P(\theta)/sin(\theta)\right]} \ ,
\label{EQ:Angle5}
\end{equation}
where the angular probability distribution for a random walk chain is given by\cite{Lasos}
\begin{equation}
P(\theta)=\frac{(1-a^2)^{3/2}\sin{\theta}}{\pi (1-a^2 \cos^2{\theta})^2}\left[\frac{1+2a^2\cos^2{\theta} \text{arccos}(-a \cos{\theta})}{\sqrt{1-a^2 \cos^2{\theta}}} +3a\cos{\theta}\right] \ ,
\label{EQ:Dangle5}
\end{equation}
with $a\rightarrow -0.25$ for long chains. 
All intramolecular sites separated more than two atoms apart interact via the pair potential of Equation \ref{EQ:Closure3}.

Initial configurations of particles were generated on a cubic lattice, and classical molecular dynamics simulations were performed using the LAMMPS MD code\cite{LAMMPS}. Due to the soft nature of the pair potential, mesoscale simulations rapidly equilibrate which is a further advantage to UA simulations of the same system. 
Simulations were run in the NVT ensemble to reproduce the conditions of the United Atom simulations.

\subsection{United Atoms Molecular Dynamics Simulations}

The above sections completely describe the coarse-grained model. It is important to note that there is no need to parameterize the model against structural correlations from any more detailed simulation (united atom or atomistic).  However, we want to demonstrate that the model can provide quantitative agreement with more detailed simulations. Therefore, in this paper we make direct comparison to united atom simulations, since the parameters of this model for polyethylene are well known. In the \textit{united representation}, each $CH_x$ group is represented as a single interaction site.\cite{UAmodel} This model has been used extensively to represent hydrocarbons of different chemical specificity.\cite{TRAPPE, Jaramillo, Jaramillo1} We ran two sets of simulations: the first set at $T=400K$ with a variety of densities and chain lengths, specifically $N=44$, $66$, $78$, $100$, $200$, each with the set of densities $\rho=0.03226$, $0.03300$, $0.03355$, $0.03441$, $0.03656$, and $0.03871$ sites/$\AA^3$. The second group of simulations was at $T=509K$ at a fixed density of $\rho=0.03153$ sites/$\AA^3$ and a variety of chain lengths, specifically $N=36$, $44$, $66$, $78$, $100$, $192$, $224$, $270$, and $1000$.

The UA simulations use the TRAPPE forcefield parameters,\cite{TRAPPE} and are performed using the LAMMPS molecular dynamics code\cite{LAMMPS}. Chains were randomly generated, and overlapping chains in the initial configuration were slowly pushed apart by a soft repulsive potential for 1 ns of simulation time. Then, the full non-bonded potential was switched on with a small time step, and the system was run for an additional 1ns while ramping up the timestep to 1.25 fs. Subsequently, chains were allowed to equilibrate for 10 ns before a final production run of 11 ns was used to collect trajectories and calculate static properties and thermodynamic averages. Although these trajectories are shorter than the longest relaxation times of long chains, here we are not concerned with dynamics; therefore these trajectories provide ample time for static properties to be collected. Simulations were run in parallel using the SDSC Trestles clusters available through the XSEDE program. All simulations were run in the NVT ensemble using Nose Hoover thermostat. 

\begin{table}[htdp]
\caption{\label{Tableparam} Polyolefin Melt Simulation Parameters}
\begin{center}
\begin{tabular}{ccclcl} \hline \hline
Lennard-Jones & $\epsilon$ (kcal/mol) & $\sigma$ ($\AA$) \\
& 0.0912 & 3.95 \\
\hline
Bond parameters & $l_0$ ($\AA$) & $k_{bond}$ (kcal/mol $\AA^2$) \\
& 1.54  & 900 \\
\hline
Angle parameters & $\theta_0$ (degrees) & $k_{angle}$ (kcal/mol rad$^2$) \\
& 114.0 & 123.75\\
\hline
Dihedral parameters & $k_1$ (kcal/mol) & $k_2$ (kcal/mol) & $k_3$ (kcal/mol) \\
& 1.4110 & -0.27084 & 3.143 \\
\hline \hline
\end{tabular}
\end{center}
\end{table}
Non-bonded interactions between sites separated by more than 4 sites away are governed by the Lennard-Jones potential,
\begin{equation}
v_{nb}(r)=4\epsilon \left[\left(\frac{r}{\sigma}\right)^{12}-\left(\frac{r}{\sigma}\right)^6\right] \ ,
\label{EQ:LJ}
\end{equation}
with the parameters specified in Table \ref{Tableparam}.
Adjacent intramolecular sites interact  through a harmonic potential 
\begin{equation}
v_{bond}(r)=\frac{k_{bond}}{2}(l-l_0)^2 \ ,
\label{EQ:Harmonic}
\end{equation}
where $l$ is the bond length. Angle interactions between triplets of intramolecular sites are governed by a harmonic potential of the form
\begin{equation}
v_{angle}(\theta)=\frac{k_{angle}}{2}(\theta-\theta_0)^2.
\label{EQ:Angle}
\end{equation}
Finally, dihedral interactions between quadruplets of atoms is governed by the OPLS dihedral potential\cite{OPLS}
\begin{equation}
v_{dih}(\phi)=\frac{1}{2}k_1[1+\cos{(\phi)}]+\frac{1}{2}k_2[1-\cos{(2\phi)}]+\frac{1}{2}k_3[1+\cos{(3\phi)}] \ .
\label{EQ:OPLS}
\end{equation}

\section{Structural Correlation Functions and Isothermal Compressibility}
The particles in the coarse grained model correspond to fictitious center of mass sites of the underlying real chain, either at the single soft sphere level, or at the mutiblock level. In other words, many atoms in the underlying detailed description are ``mapped" onto a single site in the coarse grained representation. Thus, the structure of the coarse grained model reproduces the structure of the detailed model only on the level of the mapping, and the structures are equivalent if the distribution functions are equivalent at that level of description and larger. Many atomistic configurations could have the same center of mass configuration, and a coarse grained procedure is unable to distinguish between these ``finer" levels of detail. This leads to a reduction in the dimensionality of  the configuration space which is also associated with a smoothing of the probability distributions which have a consequence for the entropy of the system, which will be discussed below. 

\begin{figure}
\includegraphics[scale=0.6]{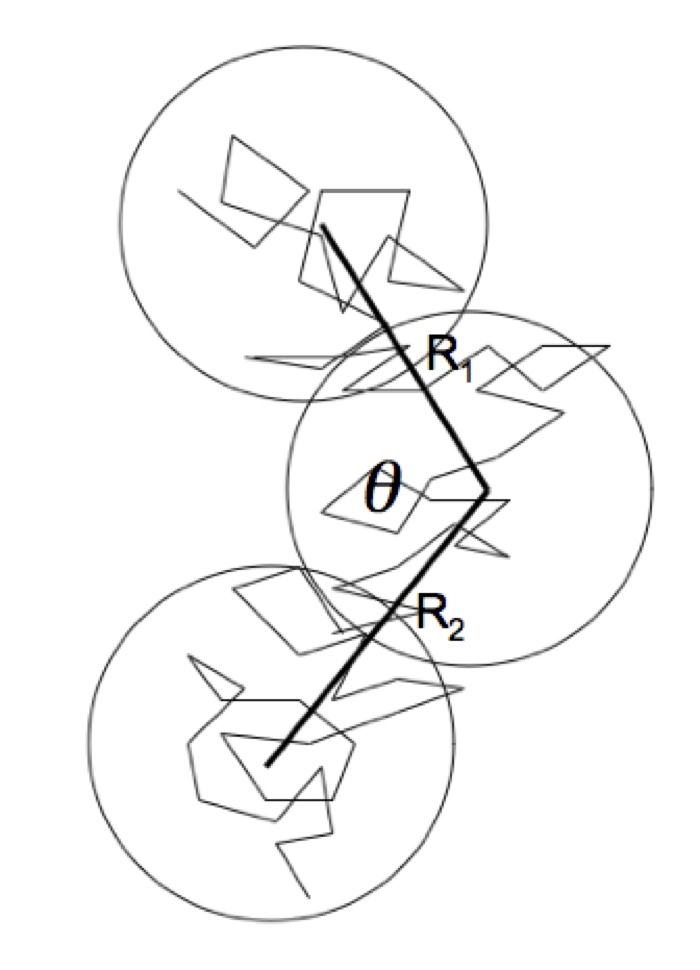}
\caption{Depiction of the three block model mapped onto a random walk. The coarse-grained coordinates are $R_1$ and $R_2$, which are the bond vectors joining effective coarse grained sites, and the angle, $\theta$, between them. Note that many underlying atomistic configurations are possible for each configuration of the coarse-grained coordinates.}
\label{FG:AngleCartoon}
\end{figure}

\begin{figure}
\includegraphics[scale=0.6]{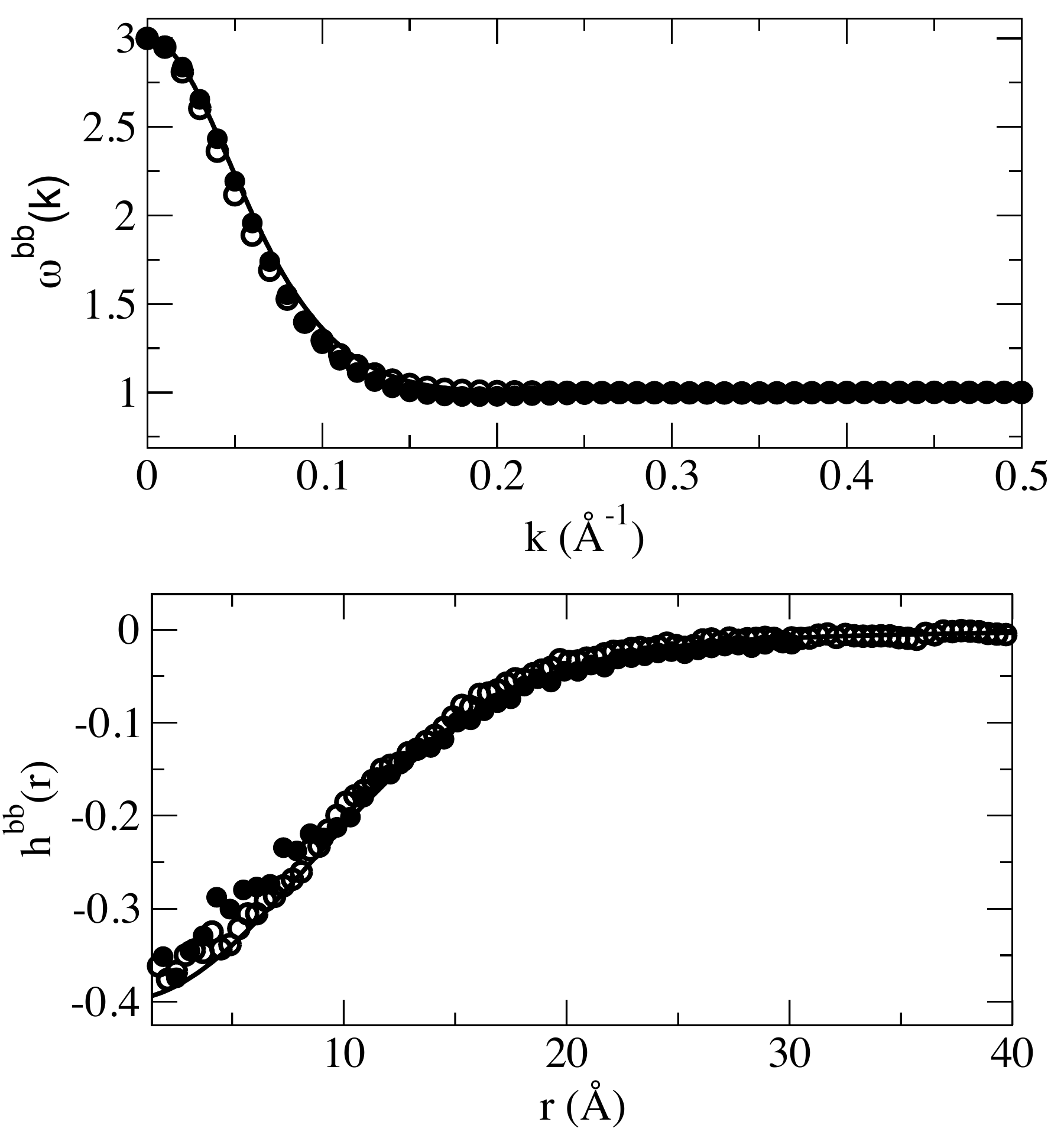}
\caption{Comparison of the intramolecular distribution, $\omega^{bb}(k)$ (top) and intermolecular correlation function, $h^{bb}(r)$, (bottom) between block centers calculated from both 3-block and united atom simulations of PE225. The solid line depicts theoretical predictions from the coarse-grained model.}
\label{FG:Distributions}
\end{figure}

Here we are interested in comparing the \textit{structural distributions} between coarse-grained and united atom simulations in the configuration space of the coarse-grained model. Figure \ref{FG:AngleCartoon} shows a depiction of a random-walk chain mapped onto a tri-block chain with three interaction sites located at the center of mass of a group of monomers. The bond vector linking coarse-grained sites is given as $\textbf{R}$ and the angle, $\theta$, defines the angle between two bond vectors, $\textbf{R}_1$ and $\textbf{R}_2$. As a representative case, we map a polyethylene chain with $N=225$ monomers onto a 3-block model, each block with 75 underlying monomers. 
%Figure \ref{FG:AngleDist} shows the distribution of the bond vector and angle vector from both coarse-grained and united atom simulations. These are compared with the equations for the distributions in our model, Equation \ref{EQ:BondGaussch5} and \ref{EQ:Dangle5}, and show good consistency between theory and simulations. The discrepancy that can be observed in the plot of the segment length distribution is likely due to numerical error, which is a consequence of the poor statistics, i.e. the small number of chains present in the atomistic simulations of large chains.
The top panel of Figure \ref{FG:Distributions} displays the intramolecular block-block correlation function in reciprocal space, for the central block in a chain represented as a tri-block chain. The distribution is calculated with the analytical expression of the tri-block model, and directly compared to the united atom simulation, and the mesoscale simulation of the coarse-grained, tri-block representation.

The bottom panel of Figure \ref{FG:Distributions} shows the intermolecular correlation function, $h^{bb}(r)$, between block centers from the analytical expression, and both the coarse-grained and united atom simulations. The agreement is quantitative, with the only parameter that enters the coarse-grained potential, $c_0$, calculated from numerical PRISM.
Typically the ratio between this distribution computed in both the coarse-grained and united atom simulation would be used in an iterative numerical scheme to optimize the pair potential or the related force. Our approach does not require any optimization since the two distributions are equivalent. This is because the coarse-grained potential reproduces the underlying free energy surface along the coarse-grained degrees of freedom and thus reproduces the correct probability distributions of coarse-grained coordinates.  All the structural properties display consistency independent of the level of coarse-graining. 

%\begin{center}
%\begin{figure}
%\includegraphics[scale=0.5]{FigureAngleDist.pdf}
%\caption{Distribution of the bond vector (top) and angle (bottom) from coarse-grained %simulations (open circles) compared to united atom simulations (filled circles). The solid line %indicates the predicted distribution for a Gaussian chain.}
%\label{FG:AngleDist}
%\end{figure}
%\end{center}

It is straightforward to show that the \textit{compressibility} is also preserved in coarse-graining. The isothermal compressibility, $\kappa_T$, is calculated from liquid state theory as
\begin{equation}
\hat{S}(k=0)=\rho k_B T \kappa_T \ ,
\label{EQ:compress}
\end{equation}
with $\hat{S}(k=0)$ the $k \rightarrow 0$ limit of the static structure factor $\hat{S}(k)$.
Taking this definition in the two levels of description, for the monomer (m) and for the blob (b) respectively, we have 
\begin{eqnarray}
\rho_m k_B T \kappa^{mm}_T -\rho_b k_B T \kappa^{bb}_T&=& \hat{S}^{mm}(0)-\hat{S}^{bb}(0) \ , \nonumber  \\
&=&[\hat{\omega}^{mm}(0)+\rho_m \hat{h}^{mm}(0)]-[\hat{\omega}^{bb}(0)+\rho_{b}\hat{h}^{bb}(0)] \ ,  \\
&=& [N+\rho_m \hat{h}^{mm}(0)]-\frac{1}{N_b}[ N+\rho_m \hat{h}^{bb}(0)] \ , \nonumber 
\end{eqnarray}
hence
\begin{eqnarray}
\rho_b k_BT \kappa_T^{bb}&=&\frac{\rho_m k_BT \kappa_T^{mm}}{N_b} \ , \nonumber \\
 \kappa_T^{bb}&=& \kappa_T^{mm}  \ ,
 \label{EQ:kappa}
 \end{eqnarray}
 for any level of block description. We have used here the property that $\hat{h}^{bb}(0)=\hat{h}^{mm}(0)$, as can be seen from Equation \ref{EQ:HbbK5} given that $\hat{\Omega}^{bm}(0)=\hat{\Omega}^{mm}(0) = 1$, as required by the normalization.
As it is expected, the compressibility of the liquid is independent of the level of detail that is used in the representation of the polymer chain. In this way, the described coarse-grained representations can be conveniently adopted in molecular dynamic simulations of polymer liquids, together with the related potentials described in the previous section: this coarse grained description allows one to select the most convenient length scale for the molecular representation, which depends on the properties that one wants to study, while being confident that both structure and compressibility are correctly represented. More thermodynamic properties are studied in the following section.

%~~~~~~~~~~~~~~~~~~~~~~~
\section{Equation of State, Free Energy, Internal Energy, and Entropy}
This section shows that not only the structure and the isothermal compressibility, but also the thermodynamic quantities of pressure, and excess free energy are conserved quantities across the variable levels of representation. In other words, the multi-scale scheme is self-consistent across multiple block levels of description and once the parameters are correctly chosen reproduces the correct equation of state as those predicted from UA simulations. We also show that other thermodynamics quantities, such as the internal energy and the entropy depend on the level of coarse-graining, as it was expected.

\subsection{Equation of State}
As a first test of the capability of the coarse-grained potential to provide thermodynamic consistent \textit{simulations} across multiple levels of representation, we performed a series of simulations of both soft-blobs with variable levels of coarse-graining and UA models under the same set of conditions. The pressure is calculated from each simulation using the standard viral expression, with $n$ the total number of monomers
\begin{equation}
P=\frac{nk_BT}{V}+\frac{\sum_i^nr_i\cdot f_i}{3V}.
\label{EQ:LAMMPSvirial}
\end{equation}

\begin{center}
\begin{figure}
\includegraphics[scale=0.6]{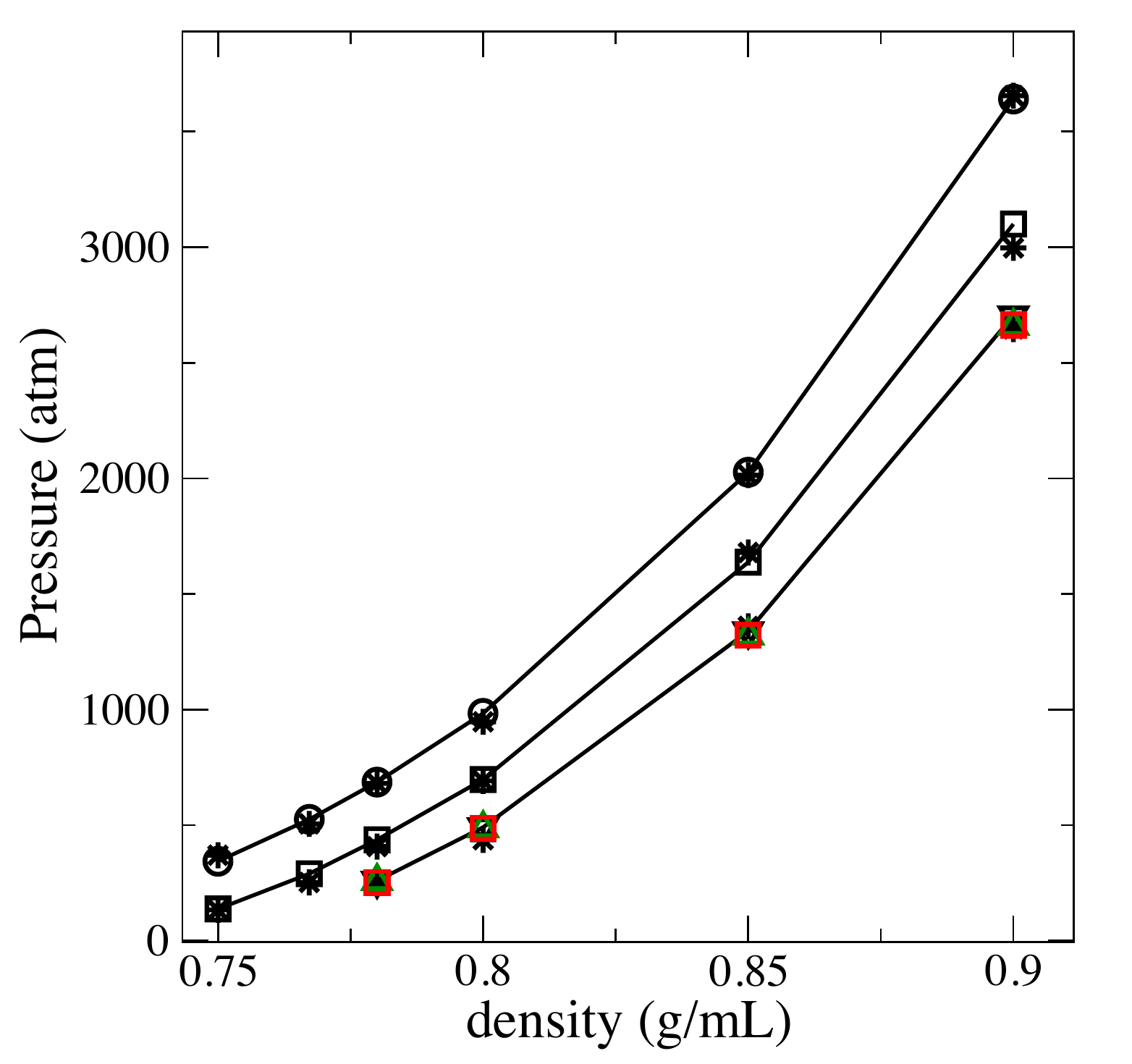}
\caption{Comparison of the pressure from united atom simulations of PE $N=44$ (black circles), PE $N=78$ (squares) and PE $N=200$ ( black triangle) with coarse-grained simulations of the soft sphere model (stars) and tri-block model (green triangles) and 5-block model (red squares). Lines are a guide to the eye.}
\label{FG:Thermo}
\end{figure}
\end{center}

\begin{center}
\begin{figure}
\includegraphics[scale=0.6]{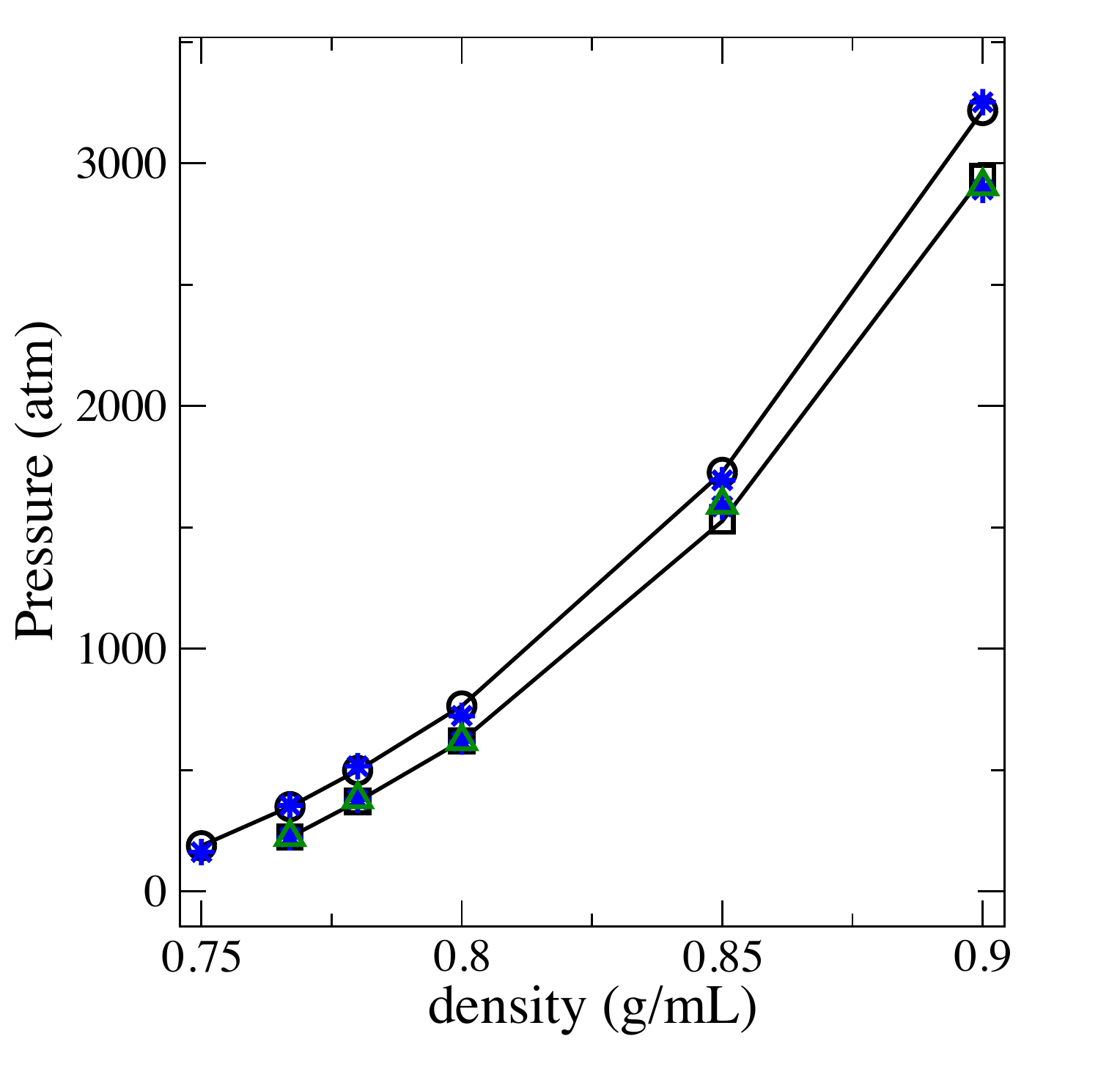}
\caption{Comparison of the pressure from united atom simulations of PE $N=66$ (black circles), PE $N=100$ (black squares) with coarse-grained simulations at the soft sphere level (stars) and 3-block level for PE100 (green triangles). Lines are a guide for the eye.}
\label{FG:Thermo2}
\end{figure}
\end{center}

\begin{center}
\begin{figure}
\includegraphics[scale=0.6]{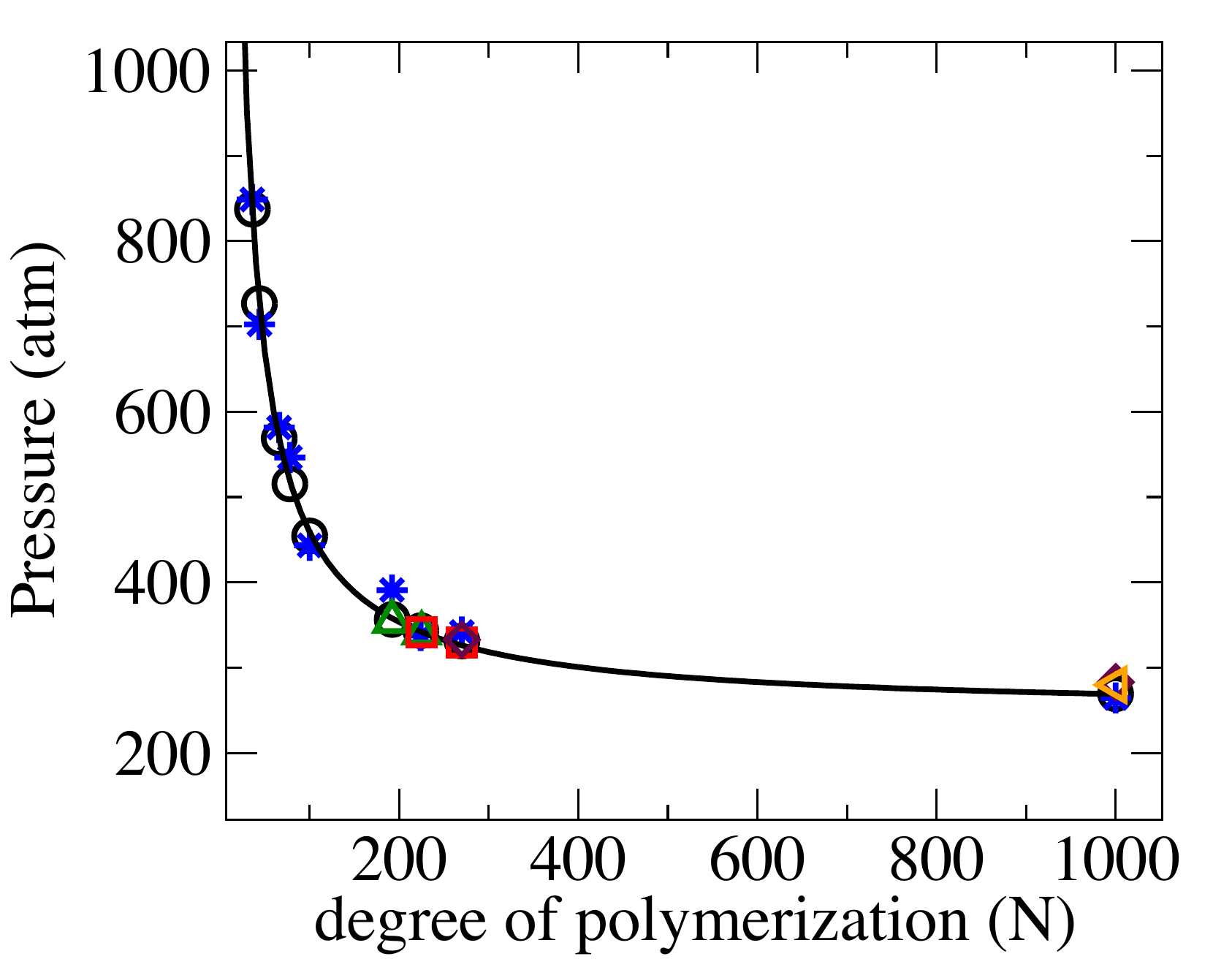}
\caption{Pressure vs. degree of polymerization calculated for a hierarchy of soft-block simulations as compared to united atom simulations. All simulations were carried out at constant temperature, T=509K, and density, $\rho=0.733$ g/mL. United atom model is depicted with black circles, soft sphere model with blue asterisk, tri-block with green triangles, 5-block with red squares, 10-block with maroon diamonds, and 20-block with orange left-oriented triangle. The line is the theoretical prediction of Eq. \ref{EQ:Soln1ch4} with $c_0$ dfined as in the right panel of Figure 5.}
\label{FG:Richter}
\end{figure}
\end{center}

Figures \ref{FG:Thermo} and \ref{FG:Thermo2} show the pressure from polyethylene simulations performed at $T=400 K$ for a variety of chain lengths and densities. In all cases the coarse-grained simulations reproduce quantitatively the trend of pressure as a function of density of the equation of state without any post-optimization scheme or fitting procedure, and across variable levels of coarse-graining. We also consider a set of simulations of increasing chain length performed at a constant monomer density of $\rho=0.733 g/ml$ and temperature, $T=509K$. As united atom simulations of long chains are difficult to equilibrate, a reasonable starting configuration for these simulations were provided to us by Mavrantzas and coworkers.\cite{Mavrantzas, Mav2} Comparison of theses simulations with various levels of coarse-graining  are presented in Figure \ref{FG:Richter} and shows consistency of the pressure for all the samples.

We now present a formal analysis of the equation of state in the variable-level coarse-grained representation. For our model the effective potential between coarse-grained units can be accurately approximated, in the mean-field limit of a polymer liquid, by an \textit{analytical expression}, which leads for the equation of state to\cite{McCarty1,AnthonyPRL}
\begin{equation}
\frac{P}{\rho_{ch}k_BT}=1-\frac{N c_0\rho}{2} \ ,
\label{EQ:Soln1ch4}
\end{equation}
where $\rho$ is the site number density, $N$ is the number of monomers per chain, $\rho_{ch}=\rho/N$ is the density of chains, and $c_0$ is the direct correlation function at zero wave vector, $\hat{c}(k=0)$, defined from liquid state theory and discussed earlier on in this paper. This equation of state holds for any level of blob description\cite{AnthonyPRL}. All of the non-ideal contributions to the pressure that arise from highly system-specific interactions are contained in this single parameter, $c_0$. As a result, the general expression of Equation \ref{EQ:Soln1ch4} is independent of the model used, but evaluation of other thermodynamic properties from the pressure requires knowledge of the state dependence of $c_0$ for the given system of interest. Previously, we used the PRISM thread model as an informative limiting case, which predicts a linear dependence of $c_0$ on the monomer density, giving the equation of state as,\cite{McCarty1}
\begin{equation}
\frac{P_{thread}}{\rho_{ch}k_BT}=1+\frac{\pi \rho \sqrt{N} \sigma^3}{6\sqrt{6}}+\frac{\pi^2\rho^2 N \sigma^6}{216} \ .
\label{EQ:ThreadEOS}
\end{equation}
However, from numerical calculations presented above, the density dependence of $c_0$ is not linear for higher densities, and the analytical result grossly underestimates the pressure for high density systems where excluded volume effects, not included in the thread model, dominate the pressure. We now consider a convenient expression for $c_0$ that allows for the calculation of the pressure and other thermodynamic properties, which can then be tested by comparison to simulation results. When introduced into Equation \ref{EQ:Soln1ch4}, this provides an expression for the equation of state for the polymer chain.

In simple liquids, the compressibility factor ($P/\rho k_B T$) is often plotted against the packing fraction, $\eta=\pi d^3 \rho/6$, where $d$ is the hard sphere diameter. Although real polyethylene chains are not simple hard sphere fluids, a similar analysis can be made by defining an effective packing fraction $\eta_e=\pi d_e^3 \rho/6$ where $d_e$ is an ``effective" hard sphere diameter and $\rho$ is the monomer density. The effective hard sphere diameter will be smaller than the real hard sphere diameter because of the overlaps between intramolecular sites, which decreases the amount of space occupied by the molecules. We assume that for polymer chains, the direct correlation function can be written as a power series in the effective packing fraction,
\begin{equation}
c_0=-\frac{4 \pi}{3}d^3\left(1+\alpha \eta_e + \beta \eta_e^2 + \gamma \eta_e^3 + ...\right) \ .
\label{EQ:Coexpand1}
\end{equation}
Substitution into Equation \ref{EQ:Soln1ch4} gives
\begin{equation}
\frac{P}{\rho_{ch}k_BT}=1+4N\eta_e \left(1+\alpha \eta_e + \beta \eta_e^2 + \gamma \eta_e^3 + ...\right) \ .
\label{EQ:Z1}
\end{equation}
Following the intuition that for a simple hard sphere fluid, the series can be written as a linear combination of only the first and second derivatives of a geometric seres, we factor out a polynomial, $f(\eta)$, such that the remaining terms are given by the binomial coefficients of the form,
\begin{equation}
\frac{P}{\rho_{ch}k_BT}=1+4N \eta_e f(\eta_e) \sum_{n=0}^{\infty} \frac{(n+2)!}{n!2!} \eta_e^n \ ,
\label{EQ:Sum}
\end{equation}
which can be summed to give,
\begin{equation}
\frac{P}{\rho_{ch}k_BT}=1+\frac{4N \eta_e f(\eta_e)}{(1-\eta_e)^3} \ .
\label{EQ:Sum2}
\end{equation}
Next, we assume $f(\eta_e)$ to be a quadratic polynomial in terms of $\eta$, such that [$f(\eta_e)=1+c1 \eta_e + c2 \eta_e^2$], giving the final result for the equation of state for the polymer, 
\begin{equation}
\frac{P}{\rho_{ch}k_BT}=\frac{1+(4N-3)\eta_e+(4Nc_1+3)\eta_e^2 +(4Nc_2-1)\eta_e^3}{(1-\eta_e)^3} \ ,
\label{EQ:EOSexpand}
\end{equation}
and the direct correlation function,
\begin{equation}
c_0=-\frac{4\pi d^3}{3}\frac{1+c_1 \eta_e +c_2 \eta_e^2}{(1-\eta_e)^3} \ .
\label{EQ:Coexpand}
\end{equation}
Equation \ref{EQ:EOSexpand} has the form of a Carnahan Starling-type expression, but with numerical pre-factors in front which reflect the chain connectivity and the fact that the real potential is not a hard sphere potential. For large N, the equation of state can be approximated as,
\begin{equation}
\frac{P}{\rho k_BT}\approx \frac{4\left(\eta +c_1 \eta^2 +c_2 \eta^3\right)}{(1-\eta)^3} \ .
\label{EQ:PlargeN}
\end{equation}

\begin{center}
\begin{figure}
\includegraphics[scale=0.6]{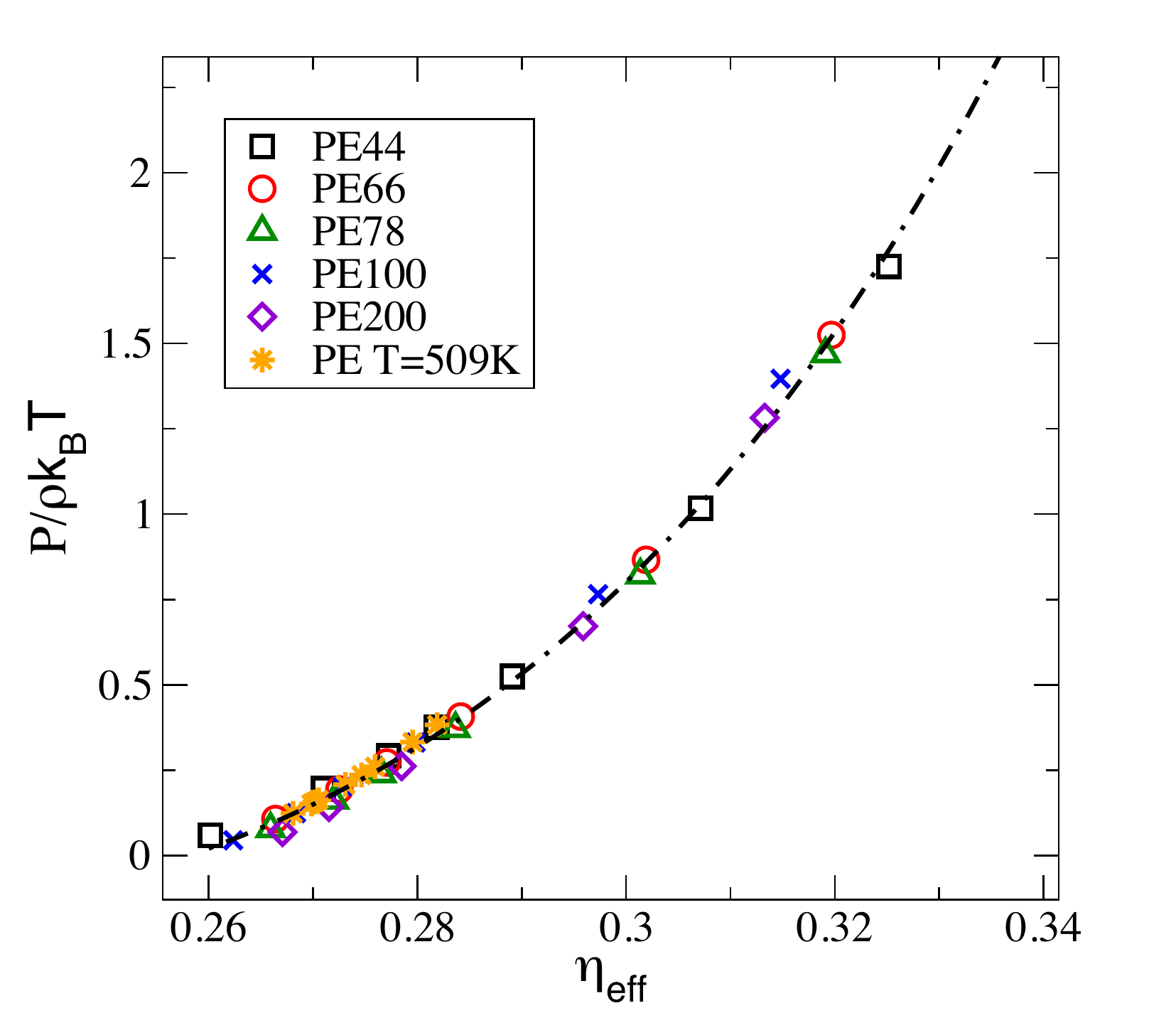}
\caption{Pressure data from United Atom simulations plotted as function of the effective packing fraction. The orange stars are the simulations at $T=509$ each with different chain length, and all the other points are simulations at $T=400$ with different chain lengths and different densities. The dot-dashed line is Equation \ref{EQ:PlargeN}, which is independent of the degree of polymerization N.}
\label{FG:Overlay}
\end{figure}
\end{center}

Assuming that $c1$ and $c2$ are independent of $N$ and temperature, a plot of the pressure against the effective packing fraction should yield a universal pressure curve for all polyethylene chains. This plot is shown in Figure \ref{FG:Overlay} where the pressure calculated from united atom simulations is presented as a function of the effective packing fraction. The points nearly fall onto a universal curve of the form of Equation \ref{EQ:PlargeN}. A fit of all the data yields the empirical values of $c1=-11.9045$ and $c2=31.1144$, which are independent of both temperature and degree of polymerization. The plot suggests that by using Equation \ref{EQ:Coexpand} we can estimate $c_0$ for any chain length at any temperature for polyethylene.

\subsection{Free Energy Estimation}
Having a form of the equation of state allows for an expression of the Helmholtz free energy change associated with a change in packing fraction from $\eta_1 \rightarrow \eta_2$, to be obtained by integration
\begin{equation}
\frac{\Delta F}{n k_B T}=\int_{\eta_1}^{\eta_2}\left(\frac{P}{\rho_{ch} k_BT}\right)\frac{d\eta'}{\eta'},
\label{EQ:FreeE}
\end{equation}
where $n$ is the number of polymer chains.
Substitution of Equation \ref{EQ:EOSexpand} into Equation \ref{EQ:FreeE} gives an expression for the free energy change per monomer
\begin{equation}
\frac{\Delta F}{Nn k_B T}=\frac{2(1-c_1-3c_2)(\eta_1^2-\eta_2^2)+4(1-c_2)(\eta_2-\eta_1)-4(c_1+2c_2)(\eta_1\eta_2^2-\eta_1^2\eta_2)}{(1-\eta_1)^2(1-\eta_2)^2}+\ln{\left[\left(\frac{1-\eta_1}{1-\eta_2}\right)^{4c_2}\left(\frac{\eta_2}{\eta_1}\right)^{1/N}\right]}
\label{EQ:FreeE2}
\end{equation}

\begin{center}
\begin{figure}
\includegraphics[scale=0.6]{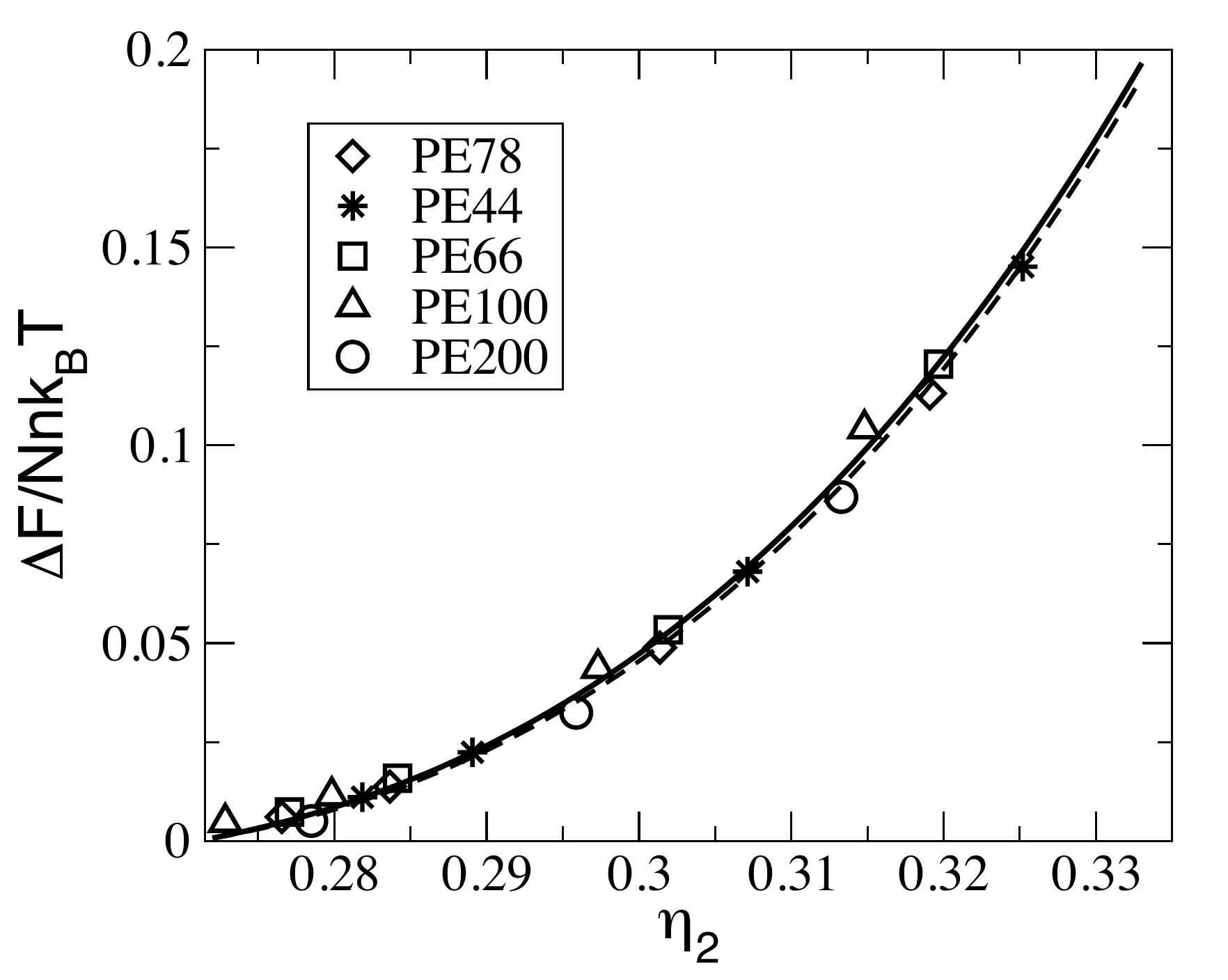}
\caption{Estimated free energy change obtained via thermodynamic integration of the pressure, as a function of packing fraction compared to a reference packing fraction of $\eta_1=0.27$, which is the lowest packing fraction that was simulated. All systems collapse to a universal curve within numerical precision, independent of degree of polymerization. The solid and dashed lines represent equation \ref{EQ:FreeE2} evaluated at N=44 and 200 respectively.}
\label{FG:FreeEnergy}
\end{figure}
\end{center}

Taking the limit as $\eta_1 \rightarrow 0$ gives a closed expression for the excess Helmholtz free energy per monomer, associated with the liquid as distinct from an ideal gas
\begin{equation}
\frac{F^{ex}}{N n k_B T}=\frac{-2(1-c_1-3c_2)\eta_2^2+4(1-c_2)\eta_2}{(1-\eta_2)^2}-4c_2\ln[(1-\eta_2)] \ .
\label{EQ:FreeE3}
\end{equation}

%\begin{center}
%\begin{figure}
%\includegraphics[scale=0.6]{FigExcessF.pdf}
%\caption{Excess free energy given by Equation \ref{EQ:FreeE3} as a function of packing fraction. For comparison, the dashed line depicts the excess free energy for a simple liquid of hard spheres.}
%\label{FG:ExcessEnergy}
%\end{figure}
%\end{center}

In the high density limit, in which we are interested, every particle interacts with a large number of surrounding molecules and the mean-field approximation applies. Since the coarse-grained potential is long ranged and bounded without short range excluded volume interactions, the excess Gibbs free energy in the canonical ensemble is calculated from the mean field approximation,\cite{Likos, Evans}
\begin{equation}
G_{exe}=\frac{n \rho_{b}}{2}  \int v^{bb}(r) g^{bb}(r) d\textbf{r} \ ,
\label{EQ:FexeC}
\end{equation}
with $\rho_b=\rho/N_b$ the blob density and $N_b$ the number of monomers per blob,
\begin{eqnarray}
\frac{G_{exe}}{nk_BT}&=&-\frac{\rho c^{bb}(k=0)}{2 N_b} \nonumber \\
&=&-\frac{\rho n_b N_b c_0}{2}=-\frac{\rho N c_0}{2} 
\label{EQ:FexeC2}
\end{eqnarray}
which is independent of the level of coarse-graining as $n_b N_b=N$, the number of monomers per chain. 
The excess Gibbs free energy per monomer is given by  $G_{exe}/(nNk_BT)=-\rho c_0/2$.   Equation \ref{EQ:FexeC2} and Equation \ref{EQ:FreeE2} are different because of the density dependence of the potential and $c_0$. Equation \ref{EQ:FreeE2} is derived by integration over the density, thus the density dependence of $c_0$ had to be accounted for. Equation \ref{EQ:FexeC2} was calculated in the canonical ensemble where the density is constant, and thus the potential is not changing. The main point here is that the excess free energy, in both ensembles, does not depend on the level of coarse-graining, and is a constant, which is consistent with Figure \ref{FG:FreeEnergy} and with the fact that the computed pressure is independent of the level of coarse graining.

\begin{center}
\begin{figure}
\includegraphics[scale=0.6]{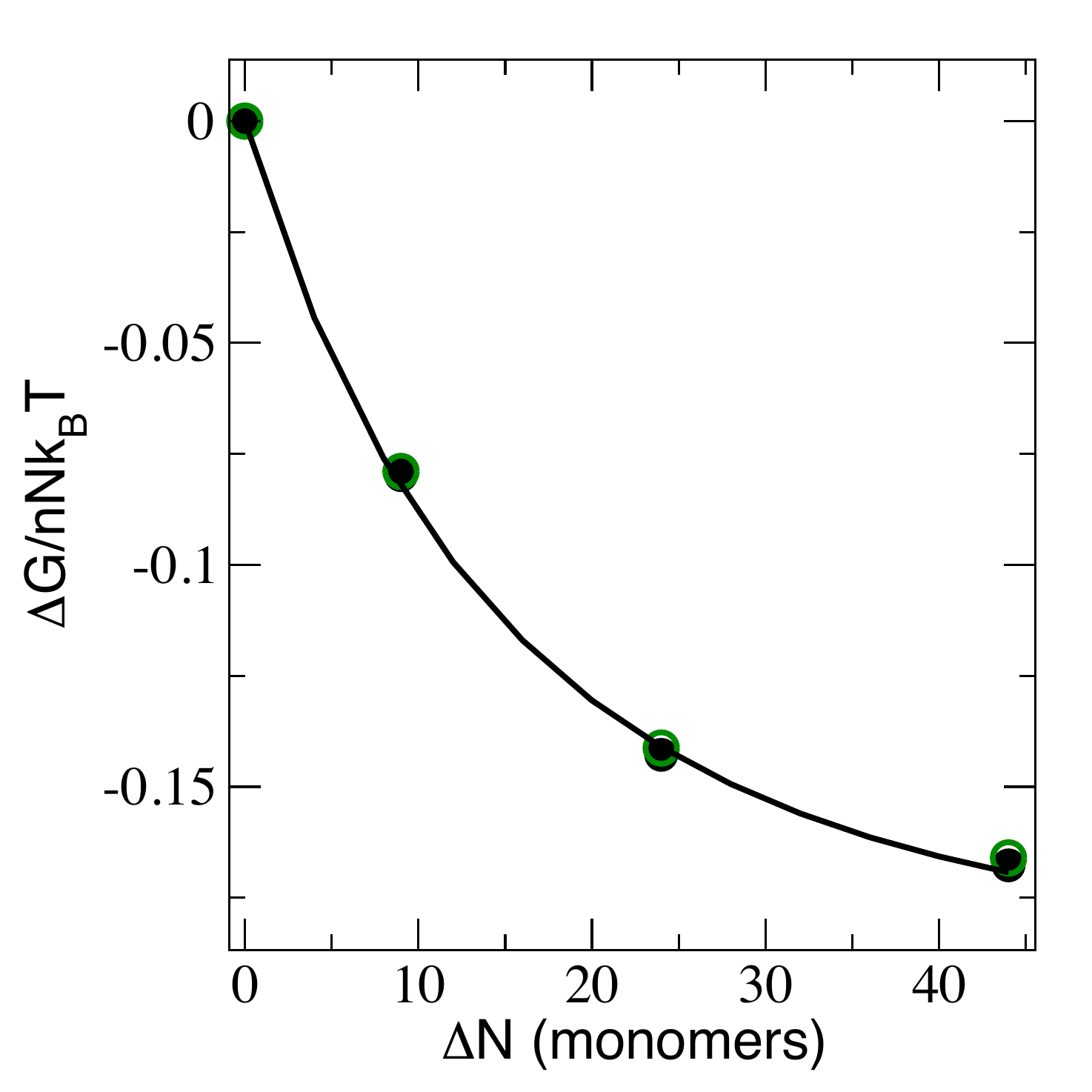}
\caption{The change in free energy per monomer associated with a change in the number of monomers per chain from a reference chain of 36 monomers as calculated from United Atom simulations (black circles) and Mesoscale simulations (open green circles) by integration of the pressure at constant volume and temperature. The solid line is the predicted free energy change from equation \ref{EQ:FEnew2}}
\label{FG:deltG}
\end{figure}
\end{center}

As an example, we performed both MS and UA simulations of short chains at constant volume and temperature ($T=400 K$ and monomer density $\rho=0.03153$ sites/$\AA^3$). Using the identity,
\begin{equation}
\frac{\Delta G}{nNk_BT}=\frac{1}{\rho k_BT}\int^{P_2}_{P_1} dP,
\end{equation}
the change in free energy was calculated numerically from the simulations, by integration of the pressure as a function of the number of monomers per chain and is shown in Figure \ref{FG:deltG}. As a test for consistency, the numerical values are compared to the predicted expression from using Equation \ref{EQ:Soln1ch4},
\begin{equation} 
\frac{\Delta G}{nNk_BT}=\left[\frac{1}{N}-\frac{\rho c_0}{2}\right]_2-\left[\frac{1}{N}-\frac{\rho c_0}{2}\right]_1 \ ,
\label{EQ:FEnew2}
\end{equation}
for which quantitative agreement between simulation data and analytical theory is observed. 

\subsection{Potential Energy in the Variable-Level Coarse-Grained Description}
We have shown so far how pressure, compressibility, structural distributions (at length-scales larger than $R_g$), and free energy of the system are equivalent between UA simulations and our coarse-grained level of representation across all levels of block description. This indicates that the coarse-graining scheme is sound since all directly observable bulk phenomena are captured. We now consider two related, but non-directly observable phenomena: the system potential energy and the entropy. 

We first examine the potential energy, which is readily calculated from the molecular dynamics simulation as the difference between the internal and the kinetic energy. 
In the coarse-grained soft sphere representation, the average potential energy is equal to the free excess energy. The entropic contributions are given by the translational entropy, which does not appear in the excess free energy, where the kinetic energy of the ideal gas is subtracted out. Thus the potential energy of the soft sphere includes only the contribution from the intermolecular pair potential, which we have previously shown to be given by
\begin{eqnarray}
\frac{E^{soft}}{nk_BT}&=&\frac{2 \pi \rho}{N k_BT}\int_0^\infty v^{cc}(r)g^{cc}(r)r^2 dr \nonumber \ , \\
                                     &\approx&\frac{-Nc_0\rho}{2} \ .
\label{EQ:Esoft}
\end{eqnarray}
Substitution of Equation \ref{EQ:Coexpand} into Equation \ref{EQ:Esoft} gives the potential energy as a function of the packing fraction, $\eta$, as
\begin{equation}
\frac{E^{soft}}{Nnk_BT}=\frac{4(\eta+c1 \eta^2+c2 \eta^3)}{(1-\eta)^3} \ ,
\label{EQ:ESoft}
\end{equation}
which, as mentioned earlier, is a free energy in the coarse-grained coordinates.

\begin{figure}
\includegraphics[scale=0.6]{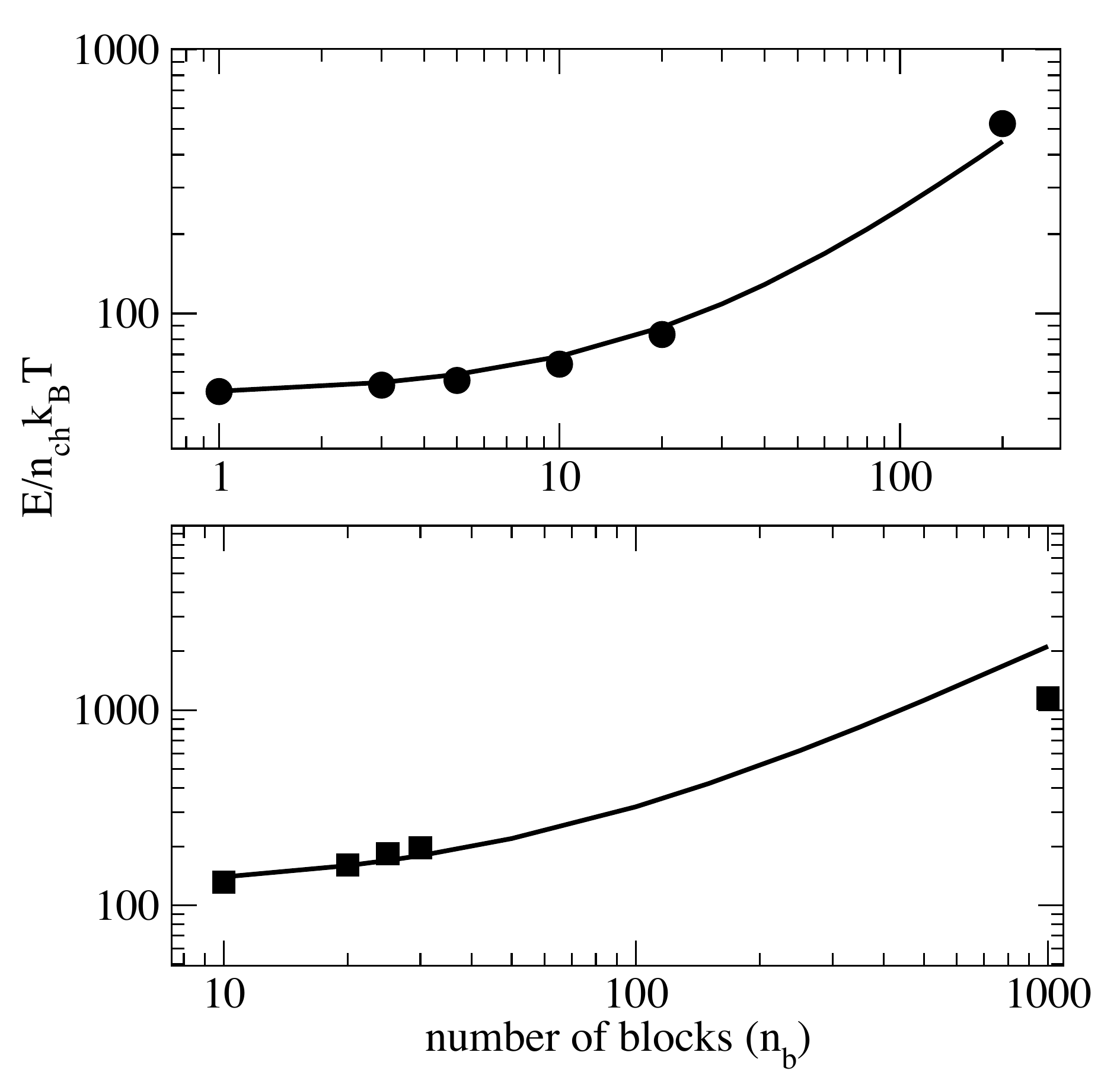}
\caption{Internal energy calculated from coarse grained simulation of PE200 (top) and PE1000 (bottom) as a function of the number of blocks ($n_b$). The last point is the internal energy from united atom simulations. The contribution from the kinetic energy has be subtracted. The solid line depicts the theoretical prediction and is extrapolated for large $n_b$.}
\label{FG:Energyplot}
\end{figure}

In the multi blob coarse-grained representation, and in the united-atom representation, the monomer-level potential energy is not represented by the excess free energy, Equation \ref{EQ:Esoft}, which corresponds to just the intermolecular contribution to the potential energy. In fact, the potential energy is a sum of intermolecular pair interactions, $U_{inter}$ and intramolecular interactions, $U_{intra}$. The intramolecular contributions can be decomposed further into separate contributions: bond stretching, angle bending, torsional rotation, and non-bonded pair interactions acting between monomers on the same chain. In this way, the potential energy depends on the level of coarse-graining and is not consistent across representations.

The potential energy in the two simulations is simply calculated as the average total internal energy minus the kinetic energy contribution. Figure \ref{FG:Energyplot} show the potential energy per molecule for two different systems, PE200 at $\rho=0.8 g/mL$ at $T=400K$ and PE1000 at $\rho=0.733 g/mL$ and $T=509K$ as a function of the number of effective sites, $n_b$. For comparison, the last data point is the potential energy from UA simulations, where the number of effective sites is equal to the number of united atoms. In both sets of simulations, the potential energy changes as the number sites is increased. 

The intermolecular component to the potential energy, due to the coarse-grained potential, is generalized from the soft sphere limit as
\begin{equation}
\frac{E^{bb}}{nk_BT}=\frac{2 \pi \rho_{b}}{k_BT}\int_0^\infty v^{bb}(r)g^{bb}(r)r^2 dr \ ,
\label{EQ:Emb}
\end{equation}
which again gives
\begin{eqnarray}
\frac{E^{bb}}{nk_BT}&=&-\frac{N \rho \hat{c}^{bb}(k=0)}{2 N_b} \ ,\nonumber \\
&=&-\frac{\rho n_b N_b c_0}{2} \ .
\label{EQ:mbintE}
\end{eqnarray}
Equation \ref{EQ:mbintE} is the internal energy due to the intermolecular coarse-grained potential, which is equivalent to the excess free energy of the system in the mean field limit. However, Figure \ref{FG:Energyplot} shows an increase in energy with the number of coarse-grained sites. This is due to the addition of intramolecular bonding and angular potentials. 
Since the bond energy is a harmonic potential with a Gaussian probability distribution, the average bond energy is simply the equipartition result, 
\begin{equation}
\left<\frac{E_{bond}}{n k_BT}\right>=\frac{3(n_b-1)}{2} \ .
\label{EQ:Ebond}
\end{equation}
Equation \ref{EQ:Ebond} is equivalent to the correction term that must be included in rescaling of the dynamics of coarse-grained simulations to account for the missing entropic degrees of freedom using a freely-rotating-chain model.\cite{IVAN1,IVAN2}
For the angular contribution to the energy we add an additional $E_{angle}\approx(n_b-2)/2k_BT$ contribution per chain. The total predicted energy, calculated from Equations \ref{EQ:mbintE}, \ref{EQ:Ebond}, and the angular contribution, is shown in Figure \ref{FG:Energyplot} with symbols representing the MS simulations and the UA simulation corresponding to the last symbol, where $n_b=N$. The line is the representation of the analytical expression for increasing number of blobs. It is clear that the energy in the coarse-grained simulations is approaching that of the united atom simulations in the limit that $n_b\rightarrow N$. 

The theoretical prediction in the long chain limit is, however, approximate because the estimated potential energy was calculated under the mean-field assumption $g^{bb}(r)\rightarrow 1$, which becomes increasingly less accurate as the local structure becomes important. Because in the limit of  $n_b \rightarrow N$, i.e. very small blocks, short range excluded volume interactions become important, and the pair distribution function $g^{mm}(r)$ presents local peaks due to packing effects, the mean-field approximation breaks down. Nonetheless, Figure \ref{FG:Energyplot} shows that  the mean field approximation does give a quite reasonable estimate of the total energy when extrapolated to the monomer limit.

\subsection{Entropy of the Variable-Lavel Coarse-Graining Model}
The basic structure of any coarse-graining procedure is the averaging of the microscopic states that are then represented by effective units, with the consequence that  the entropy of the system in a given coarse grained representation is modified with respect to its atomistic description, and that the extent of the change in entropy depends on the extent of the coarse-graining, i.e. on the level of details of the representation. In our model, which allows for the effortless tuning of the level of coarse-graining, it is possible to formalize the calculation of the entropy as a function of the level of details of the representation. As the entropy is a key quantity that needs to be correctly evaluated to properly use coarse-graining procedures, it has been the focus of several studies in relation to coarse-graining modeling. For example, we have argued that the change in entropy, together with the modification of the friction, provides the key correction contributions that need to be evaluated to properly reconstruct the correct atomistic dynamics from the accelerated dynamics measured in mesoscale simulations of coarse-grained systems.\cite{IVAN1,IVAN2}

A different type of entropy has been defined by Shell as the relative entropy, based on the information function that discriminates between coarse-grained configurations sampled in two levels of representation\cite{Shell}.  Rudzinski and Noid have discussed the relative entropy in detail in the context of numerical coarse-graining schemes such as the multiscale coarse-grained potential formalism.\cite{Noid2} The relative entropy is minimized for a coarse-grained potential that reproduces the target distribution functions, which means that the coarse-grained potential in the approach is defined as a many-body potential of mean force (not the pair-wise potential of mean force). Our coarse-grained formalism uses liquid state theory to derive a relationship between the coarse-grained sites (fictitious sites) and monomer sites. In other words, our coarse-graining procedure is devised to reproduce the correct distributions, thus minimizing the relative entropy without need for any variational approach.

The entropy of interest here is the mapping entropy, which is not related to any coarse-grained model in particular, but is an intrinsic effect of coarse-graining and a direct consequence of the fact the coarse-graining reduces the dimensionality of the configuration space and smoothes the probability distributions. The mapping entropy is related to the number of atomistic configurations which can be mapped into a single coarse-grained configuration, which can be quite large when the level of coarse-graining is extreme and the underlying chain is flexible. Noid has shown that the mapping entropy is simply the difference in entropy of the atomistic model when viewed from the atomistic configurations and the coarse-grained configurations, and adopting his notation for our coarse-graining approach the mapping entropy is given by\cite{Noid2}
\begin{equation}
S_{map}=S_\textbf{r} - S_\textbf{R}
\label{EQ:Smap}
\end{equation}
where $S_\textbf{r}$ is the entropy in the atomistic configuration, $\textbf{r} \in \{ \textbf{r}_1^1...\textbf{r}_N^{n}\}$, and $S_\textbf{R}$, is the entropy of the same set of configurations when viewed in the coarse-grained coordinates, $\textbf{R} \in \{ \textbf{R}_1^1...\textbf{R}_{n_b}^{n}\} $. In the context of this work $\textbf{R}$ is the center of mass of a group of atoms of equivalent masses, hence,
\begin{equation}
\textbf{R}_i=\sum_{j \in \text{ \ atoms  \ in \ } i} \textbf{r}_j .
\label{EQ:CM}
\end{equation}
The configurational entropy in the united atom model is 
\begin{equation}
S_\textbf{r}=-k_B \int d\textbf{r} p_{UA}(\textbf{r})\ln[V^{Nn}p_{UA}(\textbf{r})]
\label{EQ:sr}
\end{equation}
and the configurational entropy in the coarse-grained representation is 
\begin{equation}
S_\textbf{R}=-k_B \int d\textbf{R}p_{bb}(\textbf{R}) \ln[V^{n n_b} p_{bb}(\textbf{R})]
\label{EQ:sR}
\end{equation}
where $p_{UA}(\textbf{r})$ is the configurational probability distribution of the united-atom sites, and $p_{bb}(\textbf{R})$ is the probability distribution for the center of mass coordinates. V is the system volume occupied by the n molecules. Substitution of the Boltzmann weights for the probability distributions gives
\begin{equation}
S_\textbf{r} - S_\textbf{R}=\left<\frac{ E_{UA}}{T}\right>_{UA} - \left<\frac{ E_{bb}}{T}\right>_{bb}+\frac{F_{UA}}{T}-\frac{F_{bb}}{T}
\label{EQ:Sdiff1}
\end{equation}

%Since the coarse-grained simulations produce the same distributions as the underlying atomistic ones,  the coarse-grained potential is the correct pair potential that reproduces the multidimensional free energy surface along the coarse-grained coordinates. Therefore, the excess free energy is the same between coarse-grained at united atom simulations. The implications of this are that the entropy per site is simply the difference between the average energies in the two simulations and the free energy difference between ideal noninteracting chains

The excess free energy due to the effective interactions between particles was shown in the previous section to be independent of the level of coarse-graining, hence the difference in free energy in Equation \ref{EQ:Sdiff1} is due to the ideal  contribution arising from the different degrees of freedom in the two levels of representation, $F_{UA}-F_{bb}=-k_BT\ln{[Z^{UA}/Z^{bb}]}$, where $Z$ is the usual configuration integral\cite{McQuarrie}. The difference in energies in Equation \ref{EQ:Sdiff1} can be estimated by assuming both the united atom and coarse-grained chain to be ideal chains obeying Gaussian distributions, giving
\begin{eqnarray}
\frac{S_\textbf{r}}{n k_B} - \frac{S_\textbf{R}}{n k_B}&=&\frac{3}{2}(N-n_b) + \frac{3}{2}(N-n_b)+\frac{1}{2}(N-n_b)-\frac{F^{(0)}_{UA}-F^{(0)}_{bb}}{nk_BT} \nonumber \\
&=&\frac{7}{2}(N-n_b)-\ln{\left[\frac{q_{UA}}{q_{CG}}\right]}
\label{EQ:Entropy}
\end{eqnarray}
with $q^{n}=Z/V^{n}$.

As a simple approximation to the entropy associated with increasing the number of blocks in the coarse-graining procedure, we assume that the coarse-grained chain is an ideal Gaussian chain with only pair-wise interactions and harmonic bonds. The entropy is given by Equation \ref{EQ:sR} as
\begin{equation}
\frac{S_{bb}}{nk_B}\approx \frac{3}{2}n_b+\frac{3}{2}(n_b-1)-\frac{3}{2}(n_b-1) \ln{\left(\frac{3 n_b}{8\pi R_g^2}\right)}+\ln \left(\frac{Ve}{\Lambda^3 n}\right).
\label{EQ:Sbbapprox}
\end{equation}
The first two terms in Equation \ref{EQ:Sbbapprox} arise from the kinetic energy and bond potential energy, while the final two terms are the ideal translational and vibrational free energy. Importantly, there is no contribution in Equation \ref{EQ:Sbbapprox} from the potential or $c_0$, since the increasing entropy with the number of blocks $n_b$ is due solely to the increasing configurational degrees of freedom and not the interaction potential itself.

\section{Conclusion}
We have presented a systematic coarse-graining method for polyethylene chains that preserves thermodynamic consistency across multiple levels of coarse-graining without the need for any numerical optimization scheme. Instead, the procedure utilizes the PRISM Ornstein-Zernike equation to obtain the monomer-level parameter, $c_0$ which enters into the coarse-grained potential. The coarse-grained mapping procedure is formally sound, and is a result from a generalized Ornstein-Zernike equation.

 In our previous work we have parameterized the potential with the PRISM thread model to obtain a fully analytic solution of the thermodynamic properties. The analytical model is useful to understand general features of the coarse-grained potential, and quantitatively reproduces structural correlation functions. The current manuscript introduces the procedure to obtain the effective potential from numerical PRISM calculations. The procedure makes our coarse-graining method easy to implement and compatible for any system for which PRISM theory may be applied. As stated above the method is advantageous over numerically optimized potentials in that no detailed atomistic simulations need to be performed to parameterize the model. 
 
The procedure presented here has two features which are highly desirable in a coarse-graining scheme. The first is that the thermodynamic properties of pressure, excess free energy, compressibility, and structural correlations are all preserved across any level of coarse-graining. Thus, changing the level of representations does not change the state point of interest. This is important because one wishes to understand the bulk properties of a system in a well-defined thermodynamic state, and this is preserved in our coarse-graining method. The second feature is that the method allows one to easily introduce an arbitrary number of coarse-grained sites. Any coarse-graining scheme will smooth the free energy landscape, which is useful in accelerating the dynamics and sampling more configurational space. However, often one want only a small-scale increase or to be able to bridge information on different scales. Our procedure allows one to introduce systematically more and more structural details as the number of coarse-grained sites is increased. This allows us to systematically investigate how certain properties (such as the internal energy) change as a function of coarse-graining. 

In conclusion, the method here presented a detailed comparison of the coarse-grained procedure with united atom simulations of linear polyethylene. However, the method is general and widely applicable to many systems of interest.

\section{Acknowledgements}
Computational resources were provided by Trestles through the XSEDE project supported by NSF.

\end{document}